\documentclass[pre,showpacs,amsmath,amssymb,twocolumn]{revtex4-1}

\usepackage{graphicx}
\usepackage{hyperref}
\usepackage{multirow}
\usepackage{graphicx}
\usepackage{longtable}
\usepackage{answers}
\usepackage{array}

\hypersetup{breaklinks=true,colorlinks=true,citecolor=blue,linkcolor=blue,filecolor=blue,urlcolor=blue}

\begin{document}

\title{Random field disorder at an absorbing state transition in one and two dimensions}

\author{Hatem Barghathi}
\affiliation{Department of Physics, Missouri University of Science and Technology, Rolla, MO 65409, USA}

\author{Thomas Vojta}
\affiliation{Department of Physics, Missouri University of Science and Technology,
Rolla, MO 65409, USA}

\begin{abstract}
We investigate the behavior of nonequilibrium phase transitions under the influence of disorder that locally breaks the symmetry between two symmetrical macroscopic absorbing states. In equilibrium systems such ``random-field'' disorder destroys the phase transition in low dimensions by preventing spontaneous symmetry breaking. In contrast, we show here that random-field disorder fails to destroy the nonequilibrium phase transition of the one- and two-dimensional generalized contact process. Instead, it modifies the dynamics in the symmetry-broken phase. Specifically, the dynamics in the one-dimensional case is described by a Sinai walk of the domain walls between two different absorbing states. In the two-dimensional case, we map the dynamics onto that of the well studied low-temperature random-field Ising model. We also study the critical behavior of the nonequilibrium phase transition and characterize its universality class in one dimension. We support our results by large-scale Monte Carlo simulations, and we discuss the applicability of our theory to other systems.
\end{abstract}
\date{\today}

\maketitle


\section{Introduction}
\label{sec:Introduction}
 The effects of quenched disorder on phase
transitions can be drastic. For example, disorder can change the universality class of a continuous phase transition \cite{HarrisLubensky74,GrinsteinLuther76}, destroy it by smearing \cite{Vojta03a,*Vojta03b,*Vojta04}, or round a first-order phase transition
\cite{ImryWortis79,HuiBerker89,AizenmanWehr89}. In particular, disorder that \emph{locally} breaks the symmetry between two equivalent macroscopic states while preserving the symmetry globally (in the statistical sense) has strong effects on phase transitions. This type of disorder is usually called random-field disorder as it corresponds to a random external field in a magnetic
system. An experimental realization of a random-field magnet was recently found in {LiHo$_x$Y$_{1-x}$F$_4$} \cite{TGKSF06,SBBGAR07,Schechter08}; in this system, random fields arise from the interplay of dilution, dipolar interactions, and a transverse magnetic field. Moreover, impurities and vacancies generically generate random-field disorder if the order parameter of the phase transition breaks a \emph{real-space} symmetry. Such behavior occurs, e.g., in nematic liquid crystals in porous media
\cite{MCBB94} and stripe states in high-temperature superconductors \cite{CDFK06}.

Random-field disorder at equilibrium phase transitions was discussed by Imry and Ma \cite{ImryMa75}. Their argument can be summarized as follows. Consider a domain of one state embedded in a larger domain of the competing state. The formation of the domain requires a domain wall with a free energy cost of the order of the domain wall area, i.e., $L^{d-1}$ \footnote{This holds for discrete symmetry. For continuous symmetry
the surface energy behaves as $L^{d-2}$ resulting in a marginal dimension of 4}, where $L$ is the linear size of the embedded domain and $d$ is the space dimension. In contrast, the average free energy gain due to aligning the embedded domain with the prevailing local random-field is of the order of $L^{d/2}$ as follows from the central limit theorem. Consequently, in $d>2$ the system gains free energy by increasing the size of the domain without limit. On the other hand, for $d<2$, the system prefers forming domains of a limited size. Based on this heuristic argument, Aizenman and Wehr \cite{AizenmanWehr89} provided a rigorous proof that in all dimensions $d\le2$ ($d\le4$), random-field disorder prevents spontaneous symmetry breaking for discrete (continuous) symmetry. Thus, equilibrium phase transitions in sufficiently low dimensions are destroyed by random-field disorder.

Recently, nonequilibrium phase transition between different steady states have attracted lots of attention. Analogous to equilibrium phase transitions, these transitions are characterized by large-scale fluctuations and collective behavior over large distance and long times. Examples include surface growth, granular flow, chemical reactions, spreading of epidemics, population dynamics and traffic jams \cite{SchmittmannZia95,MarroDickman99,Hinrichsen00,Odor04,TauberHowardVollmayrLee05}. The effects of so-called random-mass disorder, i.e., disorder that spatially modifies the tendency toward one phase or the other without breaking any  symmetries, on nonequilibrium phase transitions have been studied in some detail. They turn out to be similar to the effects on classical and quantum equilibrium phase transitions, and include infinite-randomness criticality, Griffiths singularities, and smearing (see, e.g., Ref. \cite{Vojta06} and references therein). This similarity remains true even in the case of long-range correlated random-mass disorder \cite{IbrahimBarghathiVojta14} and for topological disorder with long-range correlations \cite{BarghathiVojta14}. Accordingly, it is important to investigate the effects of random fields on nonequilibrium phase transitions. Does an analog of the Aizenman-Wehr theorem also hold for nonequilibrium phase transitions?

To address this question, we study in this paper the generalized contact process (GCP) with two symmetric inactive states in one and two space dimensions. In the GCP, the nonequilibrium phase transition occurs between an active fluctuating phase and an inactive absorbing phase in which the system ends up in one of the inactive states, and all fluctuations cease entirely. Random-field disorder is introduced via transition rates that locally prefer one of the two competing absorbing states over the other. By studying the dynamics of the relevant degrees of freedom in the absorbing phase, which are domain walls between the two inactive states, we show that the competition between the two types of domains still ends with the system reaching one of the two absorbing states. This means that random field disorder does not destroy the absorbing state phase transition.  

\begin{figure}
\includegraphics[width=8.7cm]{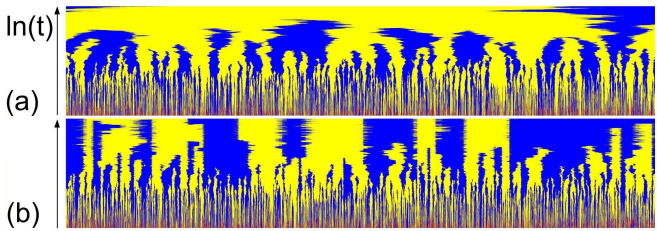}
\caption{Time evolution of the GCP in the inactive phase:
(a) without ($\mu=5/6$) and (b) with random-field disorder ($\mu_h=1, \mu_l=2/3$). 
$I_1$ and $I_2$ are shown in yellow and blue (light and dark gray). Active sites
between the domains are marked in red (midtone gray).
The difference between the diffusive domain wall motion (a) and the much slower Sinai walk (b) is
clearly visible (part of a system of $10^5$ sites for times up to $10^8$).  }
\label{fig:evo}
\end{figure}

The dynamics of the system in the inactive phase can be mapped onto that of a low-temperature random-field Ising system. In one space dimension, the long-time dynamics of the domain walls is given by a Sinai walk resulting in an ultraslow decay toward the absorbing state where the density of domain walls decays as $\ln^{-2}(t)$ (see Fig. \ref{fig:evo}). In $d\ge2$, the domain size asymptotically increases logarithmically with time. This leads to a slower decay of the domain walls density, $\ln^{-1}(t)$, than in the one-dimensional case. We also investigate the critical behavior of the phase transition between the active and inactive phases in one space dimension. At the critical point, the dynamics is even slower than in the inactive phase. We support our theoretical findings by performing large-scale Monte Carlo simulations of this model in one and two space dimensions.

This paper is organized as follows. We introduce the GCP with several absorbing states and random-field disorder in Sec. \ref{sec:Model}. In Secs. \ref{sec:Theory} and \ref{sec:Simulations} we present our theory and Monte Carlo simulation results, respectively. We conclude in Sec. \ref{sec:Conclusions}. A short account of part of this work was already published in Ref. \cite{BarghathiVojta12}. 
\section{GENERALIZED CONTACT PROCESS AND RANDOM-FIELD DISORDER}
\label{sec:Model}
First, we define the simple contact process \cite{HarrisTE74}, which is a prototypical model of a nonequilibrium phase transition. Every site $\textbf{r}$ of a $d$-dimensional hypercubic lattice can either be in the active state $A$ or in the inactive state $I$. As time evolves, inactive sites can be activated by their active nearest neighbors at a rate $\lambda m /(2d)$, where $m$ is the number of active nearest neighbors, while active sites can spontaneously become inactive at a decay rate of $\mu$. The behavior of the system is then determined by the ratio of the activation rate $\lambda$ to the decay rate $\mu$. It controls a nonequilibrium continuous phase transition between an active phase and an absorbing (inactive) phase, which is in the directed percolation (DP) \cite{GrassbergerdelaTorre79} universality class. If $\lambda\gg\mu$, the activation process survives in an infinite system for infinite times, i.e., the system reaches a steady state in which the density of active sites is nonzero, defining the active phase. In the opposite case, $\lambda\ll\mu$, all the sites in the system eventually become and remain inactive, i.e., the system will reach a state that it cannot escape, with zero density of active sites, defining the absorbing (inactive) phase. 

In the GCP introduced by Hinrichsen \cite{Hinrichsen97}, each site can be in an active state $A$ or in one of $n$ inactive states $I_k$ ($k=1\ldots n$). We define the time evolution of the GCP through the transition rates of pairs of nearest neighbors as follows:
\begin{eqnarray}
w(A A \to A I_k) = w(A A \to I_k A) &=& \bar\mu_k/n~,
\label{eq:rate_barmu}\\
w(A I_k \to I_k I_k) = w(I_k A \to I_k I_k) &=& \mu_k~,
\label{eq:rate_mu}\\
w(A I_k \to A A) = w(I_k A \to A A) &=& \lambda~,
\label{eq:rate_lambda}\\
w(I_k I_l \to I_k A) = w(I_k I_l \to A I_l) &=&
\sigma~,
\label{eq:rate_sigma}
\end{eqnarray}
with $k,l=1\ldots n$ and $k \ne l$ (all other rates are zero). For $n=1$ and $\bar \mu_k = \mu_k = \mu$, we retrieve the simple contact process with a proper rescaling of the parameters. The boundary activation rate $\sigma$ generates activity at the boundary between domains of different inactive states. This limits the number of absorbing macroscopic states to that of the inactive microscopic states $n$. In other words, the boundary activation rate $\sigma$ defined by (\ref{eq:rate_sigma}) prevents the trapping of the system in an inactive macroscopic state unless all sites are in the same inactive microscopic state. Without loss of generality, one can chose the time unit such that one of the rates equals unity, so we set $\sigma=1$. Moreover, to keep the parameter space manageable, we set $\bar\mu_k=\mu_k$ and $\lambda=\sigma=1$ \footnote{According to Ref.\ \cite{LeeVojta10}, the qualitative behavior for
          $\bar \mu_k  \ne \mu_k$ is identical to that for $\bar \mu_k = \mu_k$. Moreover, the precise value
          of $\sigma$ is not important as long as it is nonzero.},
unless otherwise mentioned. In the following, our focus will be on $n=2$  and dimensions $d=1,2$.

Consider the symmetric case, in which the decay rates toward the two inactive states $I_1$ and $I_2$ are equal, $\mu_1=\mu_2=\mu$. If $\mu$ is small enough (active phase), the system eventually reaches a steady state with nonzero density of active sites $\rho$. In this phase, the symmetry between $I_1$ and $I_2$ is not broken, since both states have identical occupation probabilities. In the opposite limit where $\mu$ is increased beyond the critical point $\mu_c^0$ ($\mu_c^0 \approx 0.628$ for $d=1$ and $\mu_c^0 \approx 1.000$ for $d=2$ \cite{Hinrichsen97,LeeVojta10,LeeVojta11}) the system undergoes a nonequilibrium phase transition to an absorbing state with all sites either in state $I_1$ or all in state $I_2$, resulting in a spontaneous breaking of the symmetry between $I_1$ and $I_2$. Therefore, the critical behavior of the transition is not in the DP universality class but in the parity conserving (PC) universality class for $d=1$ \cite{Hinrichsen97,GrassbergerKrauseTwer84,LeeVojta10} and in the generalized voter (GV) universality class for $d=2$ \cite{DCCH01,DrozFerreiraLipowski03,ACDM05,LeeVojta11}. In the asymmetric case, $\mu_1\ne\mu_2$, the favored inactive state will asymptotically play the dominant role, and the critical behavior reverts back to the DP universality class.

To introduce random-field disorder, we need to break the symmetry between $I_1$ and $I_2$ locally. Therefore, we make $\mu_1(\mathbf{r})$ and $\mu_2(\mathbf{r})$, the decay rates  at site $\mathbf{r}$ toward $I_1$ and $I_2$ respectively, independent random variables drawn from a probability distribution $W(\mu_1,\mu_2)$. A sufficient condition to preserve the symmetry globally (in the statistical sense) is $W(\mu_1,\mu_2) = W(\mu_2,\mu_1)$. Accordingly, the random variable $\alpha(\mathbf{r}) =\ln[\mu_2(\mathbf{r})/\mu_1(\mathbf{r})]$ has a symmetric probability distribution, $w(\alpha)=w(-\alpha)$. The value of $\alpha$ provides a dimensionless measure of the broken symmetry. The binary distribution
\begin{eqnarray}
W(\mu_1,\mu_2)= \frac 1 2 \delta(\mu_1 - \mu_h)\delta(\mu_2-\mu_l) \nonumber\\
+ \frac 1 2 \delta(\mu_1 - \mu_l)\delta(\mu_2-\mu_h)
\label{eq:distrib}
\end{eqnarray}
is an example, where $\mu_h$ or $\mu_l$ are the possible local decay rate values. The corresponding random variable $\alpha$ has  the symmetric probability distribution
\begin{equation}
W(\alpha) = \frac 1 2 \delta(\alpha + \alpha_0)+\frac 1 2 \delta(\alpha - \alpha_0),
\label{eq:alpha_distrib}
\end{equation}
where $\alpha_0=\ln(\mu_h/\mu_l)$.

\section{Theory}
\label{sec:Theory}
\subsection{Overview}
Let us consider the GCP in the presence of binary random-field disorder defined by (\ref{eq:distrib}). If the boundary activation process is turned off ($\sigma=0$), the difference between the two inactive states ($I_1$ , $I_2$) is no longer dynamically relevant, i.e., the system is in an inactive macroscopic state if each site is in any of the two inactive states ($I_1$ , $I_2$). In this case, the dynamics of the system is identical to that of the simple contact process with an effective decay rate $\mu_{eff} = \mu_h+\mu_l$. This results in a continuous phase transition between an active phase and an absorbing phase in which the system ends up in random combination of the states $I_1$ and $I_2$. Turning on the boundary activation rate ($\sigma>0$) favors the active phase. Moreover, the only two inactive macroscopic states are those in which all sites of the system are in the same inactive state, either $I_1$ or $I_2$ (symmetry-broken phase). In this case, the question regarding the survival of the phase transition in the presence of random-field disorder is equivalent to asking whether a symmetry-broken phase exists if $\mu_h \neq \mu_l$.

To address this question, we consider the large-$\mu$ regime where all decay rates are much larger than the clean critical value $\mu_c^0$. In this regime the decay processes (\ref{eq:rate_barmu}) and (\ref{eq:rate_mu}) dominate over the activation process (\ref{eq:rate_lambda}). In an initially active system, almost all sites quickly decay into one of the two inactive states $I_1$ and $I_2$. As a result, the system consists of a combination of domains of states $I_1$ and $I_2$. However, the domain walls can move as a result of a boundary activation process (\ref{eq:rate_sigma}) followed quickly by a decay process (\ref{eq:rate_mu}) which results in the original site being in a different inactive state. The domain wall hopping rate at site $\mathbf{r}$ thus depends on the decay rates $\mu_1(\mathbf{r})$ and $\mu_2(\mathbf{r})$ which are random. Consequently, the left-right ($d=1$) symmetry of the hopping rates is locally broken. However, their symmetry is preserved globally in a statistical sense because $W(\mu_1,\mu_2) = W(\mu_2,\mu_1)$. The resulting random walk of the domain walls with random hopping rates governs the dynamics of the system in the large-$\mu$ regime and long-time limit. 

\subsection{One space dimension, $d=1$}
A one-dimensional random walk with random hopping rates is a well-studied mathematical problem and is known as the Sinai walk \cite{Solomon75,*KestenKozlovSpitzer75,*Sinai82}. The typical displacement of a Sinai walker grows as $[\ln(t/t_0)]^{1/\psi_i}$ with time $t$ where $\psi_i=1/2$. Here, $t_0$ is a microscopic time scale, and we use a subscript $i$ on the exponent $\psi$ to mark the inactive phase. This is much slower than the $t^{1/2}$ law of the conventional random walk (see Fig. \ref{fig:evo}). When two neighboring domain walls run into each other, they annihilate, resulting in a single domain instead of three domains. The typical distance between domain walls surviving at time $t$ is therefore proportional to $[\ln(t/t_0)]^{1/\psi_i}$. Correspondingly, the density of surviving domains decays as $[\ln(t/t_0)]^{-1/\psi_i}$. As the domains grow without limit, eventually the symmetry between $I_1$ and $I_2$ will be spontaneously broken when a single domain dominates the entire system, i.e., all sites are in the same inactive state, either $I_1$ or $I_2$. The initial conditions and the details of the stochastic time evaluation of the system determine which of the two absorbing states will be the fate of the system. The existence of a symmetry broken phase implies the persistence of the nonequilibrium transition in the presence of random-field disorder. 

The time evolution of the density of active sites can also be estimated from the Sinai walk. In the large-$\mu$ regime, active sites can only exist in the vicinity of domain walls as a result of the boundary activation process. This implies that, asymptotically, the density of active sites $\rho$ is proportional to the density of the domain walls. Thus we expect that
\begin{equation}
\rho(t) \sim [\ln(t/t_0)]^{-\bar\alpha_i}~
\label{eq:rho}
\end{equation}
with $\bar\alpha_i=1/\psi_i=2$. We have introduced the decay exponent $\bar\alpha_i$ in analogy to the critical density decay exponent $\alpha$, $i$ stands for the inactive phase, as above, and the bar corresponds to a logarithmic rather than a power-law time dependence.  

To emphasize the importance of the absorbing nature of the inactive states ($I_1$, $I_2$) and its role in the survival of the nonequilibrium phase transition in the presence of a random-field disorder, we compare the domain wall dynamics in our system with that of an analogous equilibrium system, namely the random-field Ising chain. 

At sufficiently low temperatures the macroscopic state of the random-field Ising model consists of domains of up and down spins. The domain wall dynamics in the random-field Ising chain is analogous to that of our system. In fact, the hopping rates of the domain walls in the two systems can be mapped onto each other, as we show in Appendix \ref{sec:appendixA}. However, in the random-field Ising chain there is an additional process: A spin inside a domain of spin up (down) can flip down (up) due to a thermal fluctuation. This process breaks the original domain by creating two new domain walls inside it. As a result of such processes, the growth of a typical domain size is limited, preventing spontaneous symmetry breaking as suggested by the Imry-Ma criterion \cite{ImryMa75}. In contrast, in our system a site in an inactive state ($I_1$ or $I_2$) can be activated only if at least one of its nearest neighbors is in a different state (\ref{eq:rate_barmu})-(\ref{eq:rate_sigma}). As a result, the interior of an inactive uniform domain (all sites in state $I_1$ or all in state $I_2$) is dynamically dead, and the typical domain size growth is unlimited.

A more comprehensive understanding of the domain wall dynamics can be obtained from the real-space renormalization group of the random-field Ising chain developed by Fisher, Le Doussal and Monthus \cite{FisherLeDoussalMonthus98,FisherLeDoussalMonthus01,LeDoussalMonthusFisher99}. 
Translating their results into the language of the GCP, the asymptotic behavior of the linear size $R(t)$ of a domain and the density $\rho(t)$ of active sites after a quench from the active into the inactive phase (which corresponds to a decay run, i.e. a start from a completely active lattice in the Monte Carlo simulations) are found to be 
\begin{equation}
R(t) \sim [\ln(t/t_0)]^{1/\psi_i},
\label{eq:R}
\end{equation}
\begin{equation}
\rho(t) \sim [\ln(t/t_0)]^{-\bar\alpha_i}
\label{eq:N}
\end{equation}
with $\bar\alpha_i=1/\psi_i=2$. Similarly, starting from a single finite domain in the inactive state $I_1$ ($I_2$) that is embedded in an infinite system of the inactive state $I_2$ ($I_1$) (spreading runs in Monte Carlo simulations) and measuring the survival probability $P_s(t)$ of the finite domain yields  
\begin{equation}
P_s(t) \sim [\ln(t/t_0)]^{-\bar\delta_i}
\label{eq:P_s}
\end{equation}
with $\bar\delta_i=1/\psi_i-\phi=(3-\sqrt{5})/2$. Here, the linear size of the surviving domain has the same scaling behavior [Eq.(\ref{eq:R})] as the linear size $R(t)$ of a domain in the decay runs. Moreover, in the inactive phase of the GCP, active sites live only at domain walls, thus the number of active sites in a surviving system scales with the total domain wall size $R(t)^{d-1}\sim [\ln(t/t_0)]^{(d-1)/\psi_i}$.
If we define an exponent $\bar\Theta_i$ via the scaling of the number $N_s$ of active sites averaged over all systems via
\begin{equation}
N_s(t) \sim [\ln(t/t_0)]^{\bar\Theta_i}
\label{eq:N_s},
\end{equation}
then the number of active sites in surviving systems must scale as $N_s/P_s \sim [\ln(t/t_0)]^{\bar\Theta_i+\bar\delta_i}\sim [\ln(t/t_0)]^{(d-1)/\psi_i}$.
We thus obtain $\bar\Theta_i=(d-1)/\psi-\bar\delta_i$. In one dimension, $d=1$, this implies that $\bar\Theta_i=-\bar\delta_i$.

\subsection{Two space dimension, $d=2$}
\begin{figure} 
\includegraphics[width=6.0cm]{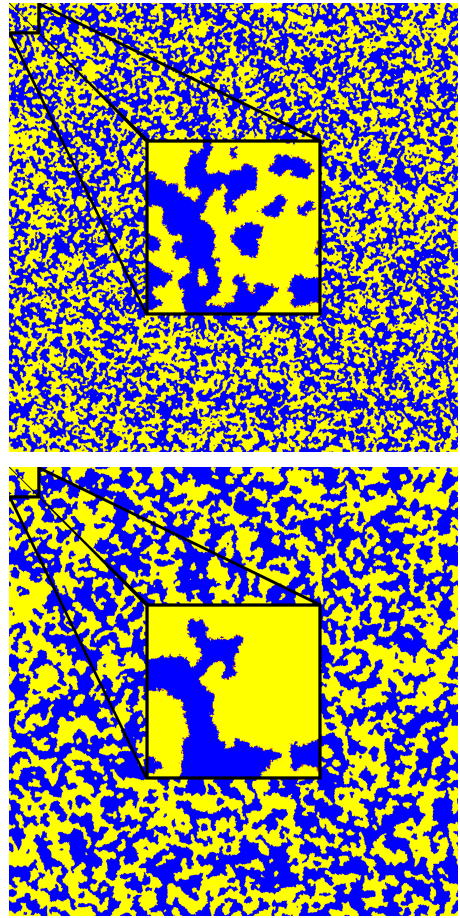}
\caption{Simulation snapshots of the two-dimensional GCP with random-field disorder, starting from a fully active lattice with size of $5000\times5000$ and $\mu\approx3.0$. $I_1$ and $I_2$ are shown in yellow and blue (light and dark gray). There is a small number of active sites at domain walls that are marked in red (midtone gray). Top: Snapshot at $t=3\times10^3$ (pre-asymptotic regime). Bottom: Snapshot at $t\approx3.6\times10^4$ (asymptotic regime).}
\label{fig:snapshot}
\end{figure}

In contrast to the one-dimensional case where the domain wall size in the inactive phase is fixed (it always consists of a single $I_1I_2$ bond); in higher space dimensions domain walls may change size (i.e., length or area; see Fig. \ref{fig:snapshot}) as the hopping of a domain wall segment might result in the annihilation of existing segments or the creation of a new ones. Therefore, the theory developed in the last section does not directly apply. However, in $d=2$, we can still map the domain wall hopping rates of the GCP with random-field disorder in the inactive phase onto those of the random-field Ising model in the low-temperature regime, as we show in Appendix \ref{sec:appendixA}. 

Grinstein and Fernandez investigated the domain growth dynamics of the random-field Ising model at low temperature following a quench from high temperature \cite{Grinstein84}. They found that the linear size $R$ of a domain grows as $\ln^2(t)$ with time up to some crossover length $R_x$, beyond which $R$ grows as $\ln(t)$. Eventually, the domain growth stops because thermal fluctuations prevent symmetry breaking in the random-field Ising model, in agreement with the Imry-Ma argument. As this mechanism does not exist in the GCP, we can ignore it. Based on their findings and the mapping (Appendix \ref{sec:appendixA}) between the GCP and the random-field Ising model, we obtain that in the inactive phase of the GCP, the linear size of a domain $R$ grows with time as 
\begin{equation}
R(t)\sim \left\{
\begin{array}{cc}\alpha_0^{-2}\ln^2(t/t_0) & (t<t_x) \\\\
                  \alpha_0^{-2}\ln(t/t_0)& (t>t_x)
\end{array},
\right. 
\label{eq:R_t_general_gcp}
\end{equation}
where $t_x$ is the crossover time between the two regimes. In contrast to the Ising model where the crossover time $t_x$ can be controlled independently by the temperature, in the GCP the ratio of $t_x/t_0$ depends only on the lattice geometry in the small-$\sigma$ limit, $\sigma\ll\mu$. (Specifically, from the Monte Carlo simulations in Sec. \ref{sec:2d-simulations} we get an estimate of $\ln(t_x/t_0)\approx 8.3$.)

In the inactive phase of the GCP, active sites exist mainly due to the boundary activation process. Therefore, active sites can only exist in the vicinity of domains boundaries. This implies that, asymptotically, the number of active sites is proportional to the total size (length) of domain walls. Accordingly the density of active sites $\rho$ is proportional to $R^{-1}$. Thus we expect that
\begin{equation}
\rho(t)\sim \left\{
\begin{array}{cc}\alpha_0^{2}\ln^{-2}(t/t_0) & (t<t_x) \\\\
                  \alpha_0^{2}\ln^{-1}(t/t_0)& (t>t_x)
\end{array}.
\right. 
\label{eq:rho_t_general_gcp}
\end{equation}

\subsection{Scaling at the critical point}
\label{sec:Scaling}
In this subsection, we give a brief summary of the scaling theory for an infinite-randomness fixed point with activated scaling. It was predicted to occur in the one-dimensional disordered contact process using a strong-disorder renormalization group \cite{HooyberghsIgloiVanderzande03} and later confirmed numerically in one, two, and three dimensions \cite{VojtaDickison05,VojtaFarquharMast09,Vojta12}. Here, we generalize it to the case where the exponents $\beta$ and $\beta^\prime $ differ from each other.

As the decay rate $\mu$ approaches its critical value $\mu_c$ starting from the active phase, the steady-state density $\rho_{stat}$ and the ultimate survival probability $P_{s}(\infty)$ approach zero, following power laws as 
\begin{equation}
\rho_{\rm stat} \sim \Delta^\beta,
\label{eq:rho_stat}
\end{equation}
\begin{equation}
P_{s}(\infty) \sim \Delta^{\beta^\prime},
\end{equation}
where $\Delta=(\mu_c-\mu_c)/\mu_c$ is the dimensionless distance from criticality, $\beta$ and $\beta^\prime$  are the order parameter and the survival probability critical exponents, respectively. Moreover, the divergence of the (spatial)
correlation length $\xi_\perp$, approaching criticality follows the power-law 
\begin{equation}
\xi_\perp \sim |\Delta|^{-\nu_\perp},
\end{equation}
where $\nu_{\perp}$ is the correlation length critical exponent. All the critical exponents defined so far describe the static behavior of observables near the critical point. The ultraslow dynamics at an infinite-randomness fixed point is reflected in the activated scaling, i.e., the correlation time $\xi_\parallel$
scales with the correlation length $\xi_\perp$ as  
\begin{equation}
\ln(\xi_\parallel/t_0) \sim \xi_\perp^\psi
\label{eq:activatedscaling},
\end{equation}
where $\psi$ is the so-called tunneling exponent and $t_0$ is a nonuniversal microscopic time scale. This leads to
\begin{equation}
\ln(\xi_\parallel/t_0) \sim |\Delta|^{-\bar\nu_\parallel}.
\end{equation}
Here $\bar\nu_\parallel=\psi\nu_\perp$ is the correlation time exponent.
Generally, the four critical exponents $\beta$, $\beta^\prime$, $\nu_\perp$ and $\bar\nu_\parallel$, form a complete set that characterizes an absorbing state phase transition. For some special cases, e.g., the transition in the DP universality class, symmetry considerations reduce this set to only three exponent, because $\beta=\beta^\prime$ \cite{Hinrichsen00}. 
In terms of these exponents we can write the finite-size (time) scaling of the density $\rho$ of active sites in a decay experiment as function of $\Delta$, $\ln(t/t_0)$, and system size $L$ as
\begin{equation}
\rho(\Delta,\ln[t/t_0],L) = b^{\beta/\nu_\perp} \rho(\Delta b^{-1/\nu_\perp},\ln[t/t_0] b^{\psi}, L b). \label{eq:rho_activated}
\end{equation}
Here $b$ is an arbitrary dimensionless scale factor. Similarly, in a spreading experiment the survival probability $P_s$, number of active sites in the active cloud $N_s$ and the mean-square radius of this cloud $R$ have the scaling forms 
\begin{equation}
P_s(\Delta,\ln[t/t_0],L) = b^{\beta^\prime/\nu_\perp} P_s(\Delta b^{-1/\nu_\perp},\ln[t/t_0] b^{\psi}, L b), \label{eq:Ps_activated}
\end{equation}
\begin{eqnarray}
N_s(\Delta,\ln[t/t_0],L) =\hspace{5.3cm}& \nonumber\\
b^{(\beta+\beta^\prime)/\nu_\perp-d}N_s(\Delta b^{-1/\nu_\perp},\ln[t/t_0] b^{\psi}, L b),\hspace{.7cm}&
\end{eqnarray}
and
\begin{equation}
 R(\Delta,\ln[t/t_0],L) = b^{-1} R(\Delta b^{-1/\nu_\perp},\ln[t/t_0] b^{\psi}, L b). \label{eq:R_activated}
\end{equation}

We can find the asymptotic time dependencies of observables in the thermodynamic limit ($L\to\infty$) and at criticality ($\Delta=0$) from the scaling relations above, by setting the scale factor $b$ to $\ln(t/t_0)^{-1/\psi}$. This leads to a logarithmic time decay of the density of active sites and the survival probability as
\begin{equation}
\rho(t) \sim [\ln(t/t_0)]^{-\bar\alpha},
\label{eq:rho_critical}
\end{equation}
\begin{equation}
P_s(t) \sim [\ln(t/t_0)]^{-\bar\delta},
\label{eq:P_s_critical}
\end{equation}
where $\bar\alpha=\beta/\bar\nu_\parallel$ and $\bar\delta=\beta^\prime/\bar\nu_\parallel$. Analogously, the number of active sites in the active cloud and the mean-square radius of this cloud starting from a single active seed site increase logarithmically with time as 
\begin{equation}
N_s(t) \sim [\ln(t/t_0)]^{\bar\Theta},
\label{eq:N_s_critical}
\end{equation}
\begin{equation}
R(t) \sim [\ln(t/t_0)]^{1/\psi}
\label{eq:R_critical}
\end{equation}
with $\bar\Theta=(d\nu_\perp-\beta-\beta^\prime)/\bar\nu_\parallel$.
This exponent relation can be rewritten in terms of the time dependence exponents as
\begin{equation}
\bar\alpha+\bar\delta+\bar\Theta=d/\psi.
\label{eq:Hyperscaling}
\end{equation}
It is similar to the hyperscaling relation for absorbing state transitions with conventional power-law scaling \cite{Hinrichsen00}.

\section{Monte carlo Simulations}
\label{sec:Simulations}
\subsection{Method and overview}
To test our predictions, we perform Monte Carlo simulations \cite{LeeVojta10,LeeVojta11} of the GCP defined by (\ref{eq:rate_barmu}) to (\ref{eq:rate_sigma}) in the presence of random-field disorder in one and two space dimensions. In the one-dimensional case, we perform the simulations with two different types of initial conditions: (i) decay runs and (ii) spreading runs. Decay runs start from a completely active lattice (all sites in state A), and we monitor the time evolution of the density $\rho$ of active sites as well as the densities $\rho_1$ and $\rho_2$ of inactive sites $I_1$ and $I_2$, respectively. Spreading runs start from a fully inactive lattice with all sites in inactive state $I_1$ except a single active (seed) site in the active state $A$. Here, we measure the survival probability $P_s$, the number of active sites in the active cloud $N_s$ and the mean-square radius $R^2$ of this cloud as functions of time. In the two-dimensional case we perform decay run simulations only. We implement the random-field disorder through the distribution (\ref{eq:distrib}) using $3\mu_l/2=\mu_h\equiv\mu$.

In both types of runs, the simulation proceeds as a sequence of individual events. Each event consists of randomly selecting a pair of nearest-neighbor sites from the active region. In the spreading runs, the active region initially consists of the seed site and its nearest-neighbors. Its size increases as activity spreads in the system. In contrast, in the decay runs, the active region is the entire system. The selected pair is updated through one of the possible processes (\ref{eq:rate_barmu}) to (\ref{eq:rate_sigma}) with probability $\tau w$. The time step $\tau$ is fixed at a constant value which is chosen such that the total probability of an outcome of the process (\ref{eq:rate_barmu})-(\ref{eq:rate_sigma}) with the highest total rate is unity. Each event result in a time increment of $\tau/N_{pair}$ where $N_{pair}$ is the number of nearest-neighbor pairs in the active region. 

\subsection{Absorbing phase in one space dimension, $d=1$}
\begin{figure}
\includegraphics[width=8.7cm]{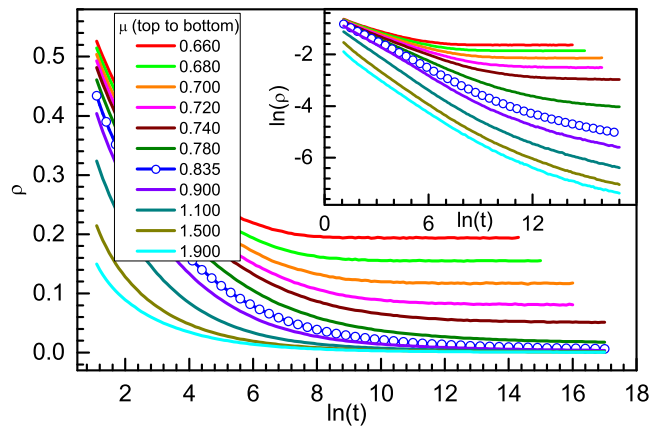}
\caption{Density $\rho$ vs time $t$ in one dimension for several values of the decay rate $\mu$. The data are averages over 60 to 1000 disorder configurations. Inset: The log-log plot shows that the density decay is slower than a power law for all $\mu$.}
\label{fig:overview}
\end{figure}

\begin{figure}
\includegraphics[width=8.7cm]{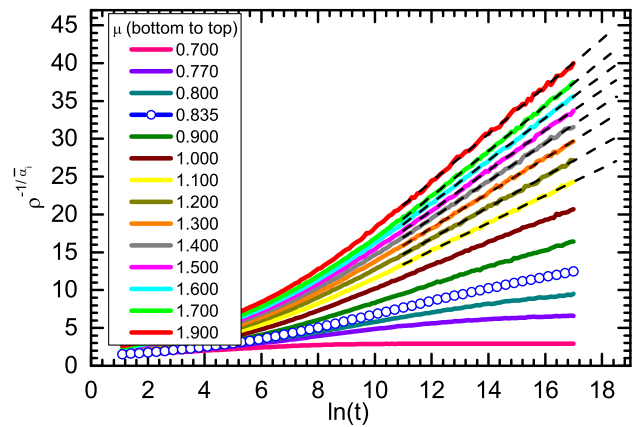}
\caption{$\rho^{-1/\bar\alpha_i}$ vs $\ln(t)$ for several values of the decay rate $\mu$. The dashed straight lines
        are fits to the predicted behavior $\rho \sim [\ln(t/t_0)]^{-\bar\alpha_i}$ with $\bar\alpha_i=2$.}
\label{fig:sinai}
\end{figure}

We studied systems with sizes up to $L=10^5$ and times up to $t_{\rm{max}}=2\times10^8$.
An overview over the density decay runs is provided in Fig. \ref{fig:overview} which shows the time evolution of the density of active sites. The inset of Fig. \ref{fig:overview} 
shows that for systems with both decay rates $\mu_h=\mu$ and $\mu_l=2\mu/3$ greater than the clean critical value $\mu_c^0=0.628$, the density $\rho$ continues to decay up to the longest times studied. Still, the decay is obviously slower than a power law.
Our theoretical arguments led to Eq. (\ref{eq:rho}), which predicts that asymptotically  $\rho^{-1/\bar\alpha_i}$  depends linearly on $\ln(t)$. This prediction is tested in Fig. \ref{fig:sinai}.
We see that all curves with $\mu>1$ follow the predicted behavior over several orders of magnitude in time.

\begin{figure}
\includegraphics[width=8.7cm]{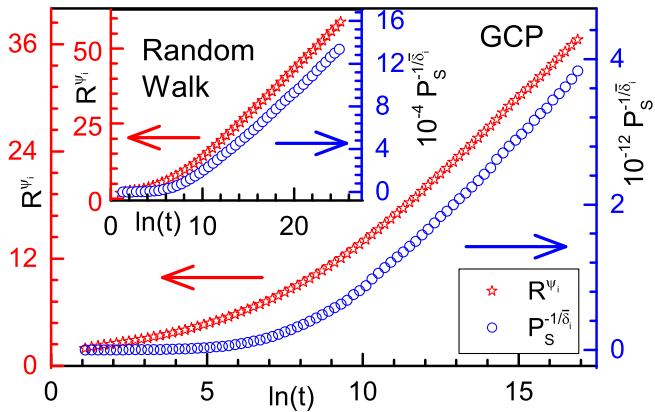}
\caption{$P_s^{-1/\bar\delta_i}$ and $R^{\psi_i}$ vs $\ln(t)$, with $\bar\delta_i=(3-\sqrt{5})/2$ and $\psi_i=1/2$. Main panel: GCP with random-field disorder, with $\mu=3$ and $\lambda=0.01$. The data are averages over $36000$ samples with $4000$ individual runs per sample. Inset: Toy model consisting of two random walkers with random hopping probabilities. The ratio between right and left hopping probabilities $\alpha_i$ at site $i$ is drawn from a time-independent binary distribution with possible values of $(2/3)^{\pm1}$. If the first walker see a ratio $\alpha_i$, the second walker sees the inverted ratio $\alpha_i^{-1}$. The data are averages over $600$ samples with $1000$ individual runs per sample.}
\label{fig:PsvzlntandRvzint}
\end{figure}

Similarly, Eqs.(\ref{eq:R}) and (\ref{eq:P_s}) predict linear dependences of both $P_s^{-1/\bar\delta_i}$ and $R^{\psi_i}$ on $\ln(t)$ (asymptotically for $t\to\infty$). To verify these predictions, we performed spreading simulations deep in the inactive phase with $\mu=3$ and $\lambda=0.01$ ($\mu/\lambda\gg1$). Our simulation results are presented in Fig. \ref{fig:PsvzlntandRvzint}.
The figure shows that $R^{\psi_i}$ meets the prediction over about one and half orders of magnitude in time. The behavior of  $P_s^{-1/\bar\delta_i}$ seems to be pre-asymptotic, i.e., $P_s^{-1/\bar\delta_i}$ slowly approaches the predicted asymptotic linear dependence on $\ln(t)$ but has not quite reached it at the end of our simulations. Increasing the time in order to reach the true asymptotic behavior, requires prohibitively large numerical effort.

As domain walls between the two inactive states are the only relevant degrees of freedom in the absorbing phase,  we used a toy model in which we replace the two domain walls ($I_1I_2$ and $I_2I_1$) in the spreading simulation by two random walkers with random right and left hopping probabilities. The ratio between right and left hopping probabilities at a given site is proportional to the ratio between the decay rates toward the two inactive states ($\mu_1/\mu_2$ for the walker representing $I_1I_2$ and $\mu_2/\mu_1$ for the other walker representing $I_2I_1$). This toy model is numerically simpler and allows us to reach longer times. The inset of Fig. \ref{fig:PsvzlntandRvzint} shows that the data for $P_s$ and $R$ obtained from the random walk toy model follow the predictions of Eqs. (\ref{eq:R}) and (\ref{eq:P_s}) over several orders of magnitude in time.

\subsection{Criticality in one space dimension, $d=1$}
We now turn to the critical point in one space dimension. In a previous work \cite{BarghathiVojta12}, we obtained a rough estimate of the critical decay rate $\mu_c$. The more detailed simulations reported here have led to a better estimate of $\mu_c$ as well as a complete set of critical exponents.

Because the critical point separates an active system from an ultimately dead one (in the absorbing state), the dynamics at criticality is expected to be slower than the dynamics in the inactive phase. Since, observables in the inactive phase evolve as power laws of $\ln(t/t_0)$, a simple power law dependent on $t$ time evolution at criticality can be ruled out. Instead, let us assume that the critical behavior follows the activated scaling scenario outlined in Sec. \ref{sec:Scaling}.

In simulations of absorbing state transitions, the critical point is often identified by plotting the data such that the critical time dependence leads to a straight line. In the case of activated scaling, this is hampered by the unknown microscopic scale $t_0$ which acts as a strong correction to scaling.

\begin{figure}
\includegraphics[width=8.7cm]{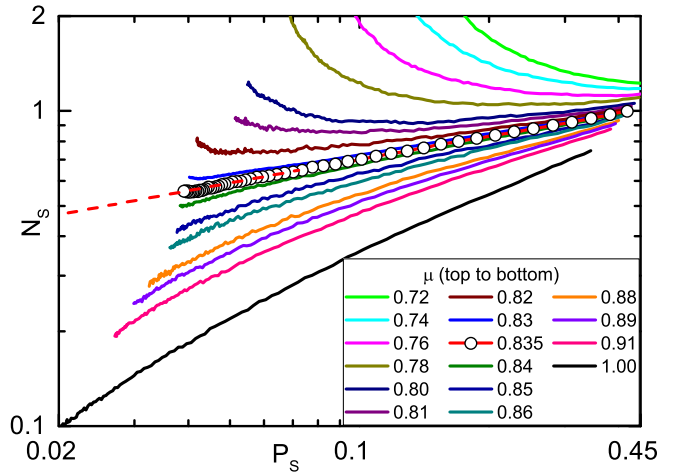}
\caption{Double-log plot of $N_s$ vs $P_s$ for several values of the decay rate $\mu$. The data are averages over 1000 to 8000 disorder configurations with 100 to 400 trials each. The straight dashed line is a power-law fit of the asymptotic part of the critical curve ($\mu=0.835$) yielding $\bar\theta/\bar\delta=-0.27(5)$.}
\label{fig:critical1}
\end{figure}

\begin{figure}
\includegraphics[width=8.7cm]{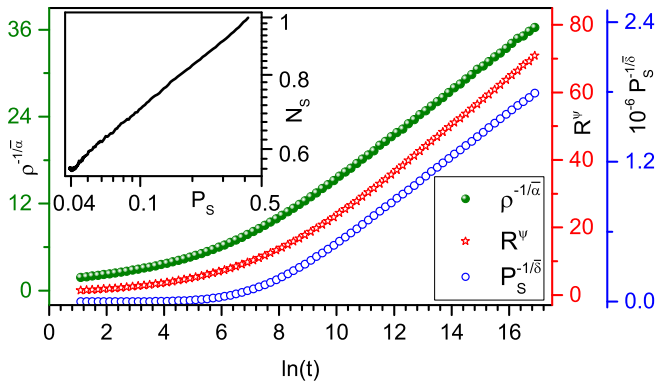}
\caption{$\rho^{-1/\bar\alpha}$, $R^{\psi}$, and $P_s^{-1/\bar\delta}$ vs $\ln(t)$ at criticality. Here, $\psi=0.62(7)$, $\bar\alpha=1.4(1)$ and $\bar\delta=0.225(8)$ are determined from the data by requiring that the corresponding curves become straight lines asymptotically. Inset: Double-log plot of $N_s$ vs $P_s$ at criticality as in Fig. \ref{fig:critical1}.}
\label{fig:critical2}
\end{figure}

However, Vojta \emph{et al}. \cite{VojtaFarquharMast09} provided a method to overcome the absence of a $t_0$ value by observing that $t_0$ should be the same for all observables measured in the same simulation run because $t_0$ is related to the time scale of the underlying strong-disorder renormalization group. Therefore, asymptotically, observables have power-law dependencies on each other. For example, combining Eqs. (\ref{eq:P_s_critical}) and (\ref{eq:N_s_critical}) gives $N_s \sim P_s^{-\bar\Theta/\bar\delta}.$ Using this method, the data plotted in Fig. \ref{fig:critical1}
indicate a critical decay rate of $\mu_c=0.835(3)$ and yield a value of $\bar\Theta/\bar\delta=-0.27(5)$. The numbers in parentheses give the error estimate of the last digits. Our error estimate contains the statistical and the systematic errors as well as the error due to the uncertainty of $\mu_c$. (Possible correlations between errors from different sources have been ignored.) To obtain the exponents $\bar\alpha$, $\bar\delta$ and $\psi$, we search for values that yield linear dependencies of each of $\rho^{-1/\bar\alpha}$, $P_s^{-1/\bar\delta}$ and $R^\psi$ [see Eqs. (\ref{eq:rho_critical}), (\ref{eq:P_s_critical}), and (\ref{eq:R_critical})] on $\ln(t/t_0)$ at the critical $\mu_c=0.835$. We find values of $\bar\alpha=1.4(1)$, $\bar\delta=0.225(8)$ and $\psi=0.62(7)$ (Fig. \ref{fig:critical2}).
Moreover, using the measured values of $\bar\Theta/\bar\delta$ and $\bar\delta$, we find  $\bar\Theta=-0.060(12)$. The hyperscaling relation, Eq. (\ref{eq:Hyperscaling}), is satisfied by the obtained critical exponents $\bar\alpha$, $\bar\delta$, $\psi$, and $\bar\Theta$, within the given errors.

\begin{figure}
\includegraphics[width=8.7cm]{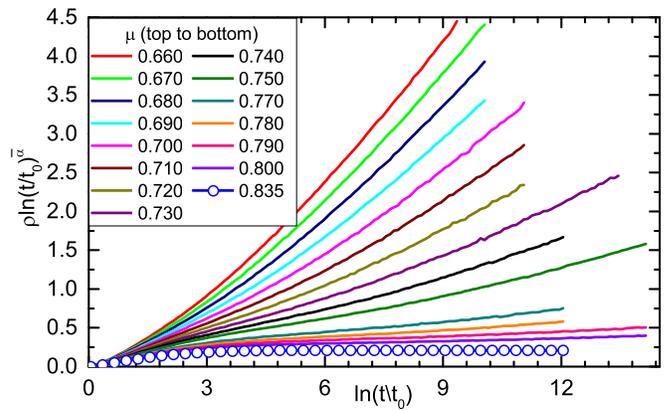}
\caption{$\rho\ln(t/t_0)^{\bar\alpha}$ vs $\ln(t/t_0)$ for several decay rates $\mu$ at and below the critical decay rate $\mu_c=0.835$. The quantity $\rho\ln(t/t_0)^{\bar\alpha}$ has zero scale dimension. Thus, asymptotically it is time independent at criticality, $\mu_c=0.835$.}
\label{fig:scaling1}
\end{figure}

So far, we obtained only three independent critical exponents. In order to find a complete set of critical exponents that is required to characterize the universality class of the transition, we need to find one more critical exponent independently. Thus, we turn to the density scaling relation, Eq. (\ref{eq:rho_activated}). Setting the scale factor $b=\ln(t/t_0)^{-1/\psi}$ and in the limit $L\to\infty$, we get
\begin{equation}
\rho\ln(t/t_0)^{\bar\alpha} =\tilde{X}[\Delta^{\bar\nu_\parallel}\ln(t/t_0)]. \label{eq:x_scaling}
\end{equation}
Here $\tilde{X}$ is a scaling function. At the critical point ($\Delta=0$), the quantity $\rho\ln(t/t_0)^{\bar\alpha}$ asymptotically approaches a constant value [$\tilde{X}(0)=$ const.]. As we deviate from the critical point toward the active phase, the quantity $\rho\ln(t/t_0)^{\bar\alpha}$ represents the scaling function $\tilde{X}$ with an argument that is scaled by $\Delta^{\bar\nu_\parallel}$ as shown in Fig. \ref{fig:scaling1}.

\begin{figure}
\includegraphics[width=8.7cm]{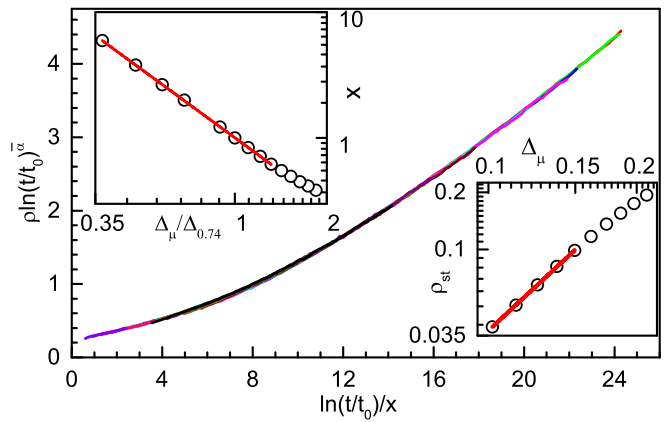}
\caption{Scaling plot of $\rho\ln(t/t_0)^{\bar\alpha}$ vs $\ln(t/t_0)/x$ for several decay rates $\mu$ below the critical decay rate $\mu_c=0.835$ (the same off-critical decay rate values listed in Fig. \ref{fig:scaling1}). $x$ is the scaling factor necessary to scale the data onto the curve of $\mu=\mu_{ref}=0.74$.  Upper inset: Double-log plot of the scaling factor $x$ vs $\Delta_\mu/\Delta_{0.74}$ where $\Delta_\mu=(\mu_c-\mu)/\mu_c$. The straight solid line is a power-law fit yielding $\bar\nu_\parallel=1.78(4)$. Lower inset: Double-log plot of the stationary density $\rho_{st}$ vs $\Delta_\mu$. The straight solid line is a power-law fit yielding $\beta=2.42(8)$.}
\label{fig:scaling2log}
\end{figure}

In Fig. \ref{fig:scaling2log}, we rescale the abscissa of each of the off-critical curves with a scaling factor $x$ until they all collapse onto a reference curve.
According to Eq. (\ref{eq:x_scaling}), a fit of the scaling factor $x$ to the power-law dependence $x=(\Delta/\Delta_{ref})^{-\bar\nu_\parallel}$ (see upper inset in Fig. \ref{fig:scaling2log}), yields the correlation time critical exponent $\bar\nu_\parallel=1.78(4)$.

With the help of the scaling relations [Eqs. (\ref{eq:rho_activated})-(\ref{eq:R_activated})], other critical exponents can be calculated (Table \ref{tab:exponents}),
\begin{table}
\caption{Critical and inactive phase exponents for the one dimensional generalized contact process with two symmetric inactive states in the presence random-field disorder. The values for the inactive phase are found analytically. The values for the generic transition emerge from fits of our data (above the horizontal line) and from scaling relations (below the horizontal line). The numbers in parentheses gives the estimated error of the last given digits, where possible correlations between errors from different  sources are ignored.}
\centering  
\renewcommand*{\arraystretch}{1.25}
\begin{tabular*}{8cm}{@{\extracolsep{\fill}}c|c||c|c}
\hline\hline
  \multicolumn{2}{c||}{\textbf{Critical point}} & \multicolumn{2}{c}{\textbf{Inactive phase}}\\
\hline
\centering  
 $~~~~~~\bar\alpha~~~~~~$  & 1.4(1)    &  $~\bar\alpha_i~~~~~$  &    2       \\
 $\psi$                    &  0.62(7) &  $~\psi_i~~~~~$             &    1/2     \\
 $\bar\delta$              & 0.225(8)&  $~\bar\delta_i~~~~~$       &   $(3-\sqrt{5})/2$  \\
 $\bar\nu_{\parallel}$              & 1.78(4)&  &   \\
 $\beta$              & 2.42(8)&  & \\ 
  \cline{1-2}
$\bar\Theta$              & -0.060(12)&         $~\bar\Theta_i~~~~~$    & $(\sqrt{5}-3)/2$   \\
 $\nu_\perp$              & 2.9(4)&  & \\ 

 $\beta^\prime$              & 0.40(2)& & \\   
\hline\hline
\end{tabular*}
\label{tab:exponents}
\end{table}
e.g., the scaling relation $\beta=\bar\alpha\bar\nu_\parallel$ gives the order parameter critical exponent $\beta=2.5(2)$. The steady-state density $\rho_{stat}$ [Eq. (\ref{eq:rho_stat})] yields another independent estimate of the exponent $\beta=2.42(8)$
(see lower inset in Fig. \ref{fig:scaling2log}). 

We conclude that all the Monte Carlo simulations data are well described within the activated scaling scenario.

\subsection{Two space dimension, $d=2$}
\label{sec:2d-simulations}
\begin{figure} 
\includegraphics[width=8.7cm]{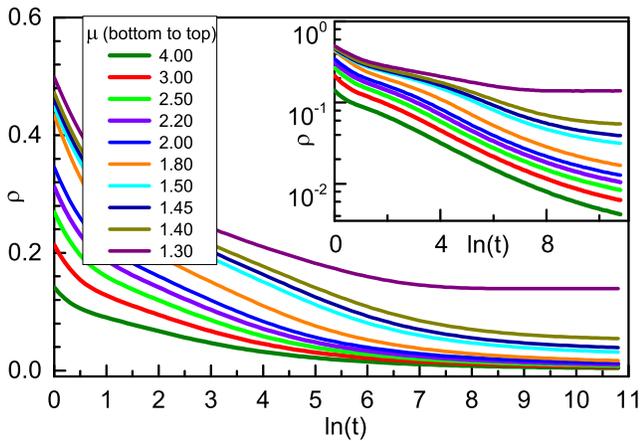}
\caption{Density $\rho$ vs time $t$ in two dimensions for several values of the decay rate $\mu$. The data are averages over 100 disorder configurations. Inset: The log-log plot shows that the density decay is slower than a power law for all $\mu$.}
\label{fig:RhoVisTime}
\end{figure}

In two dimensions our simulations focused on the inactive phase. We studied systems with sizes of up to $2000\times2000$ sites and times up to $t_{\rm{max}}=5\times10^4$. Figure \ref{fig:RhoVisTime}
shows an overview of the time evolution of the density of active sites from decay runs.
 Similar to the one-dimensional case, in two-dimensional systems with both decay rates $\mu_h=\mu$ and $\mu_l=2\mu/3$ greater than the clean critical value $\mu_c^0=1.000$, the density $\rho$ continues to decay slowly (slower than a power law) up to the longest times studied, as shown in the inset of Fig. \ref{fig:RhoVisTime}.

\begin{figure} 
 \includegraphics[width=8.7cm]{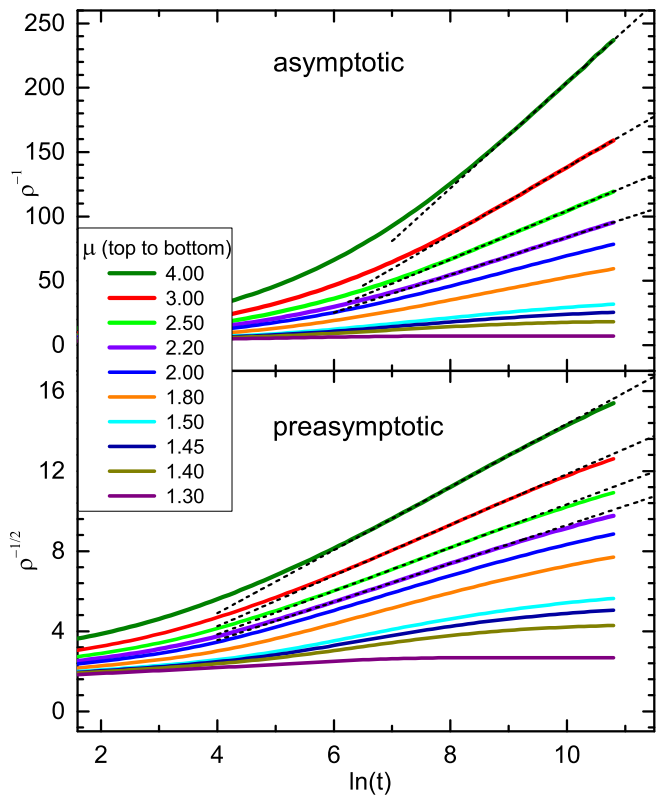}
 \caption{Upper panel: $\rho^{-1}$ vs time $t$ for several values of the decay rate $\mu$. The solid straight lines are fits to the predicted asymptotic behavior $\rho \sim \ln^{-1}(t/t_0)$. Lower panel: $\rho^{-1/2}$ vs time $t$ for several values of the decay rate $\mu$. The solid straight lines are fits to the predicted pre-asymptotic behavior $\rho \sim \ln^{-2}(t/t_0)$.}
 \label{fig:2in1InvRhoVisTime}
 \end{figure}

 According to our theory [Eq. (\ref{eq:rho_t_general_gcp})] the time evolution of the density of active sites $\rho$ is predicted to consist of two regimes, a pre-asymptotic regime and an asymptotic regime. In the pre-asymptotic regime $\rho^{-1/2}$ depends linearly on $\ln(t)$ up to a crossover time $t_x$, after which $\rho^{-1}$ depends linearly on $\ln(t)$. The prediction of Eq. (\ref{eq:rho_t_general_gcp}) is tested in Fig. \ref{fig:2in1InvRhoVisTime}, 
  which it shows that for all curves with $\mu>2$, the predicted behavior is evident up to the longest times studied. Moreover, our results give an estimate of $\ln(t_x/t_0) \approx 8.3$. 
\section{CONCLUSIONS}
\label{sec:Conclusions}
To summarize, we have studied the effects of random-field disorder on the nonequilibrium phase transitions of the one- and two-dimensional GCP. We have found that these transitions survive the presence of random-field disorder, in contrast to equilibrium transitions in one and two space dimensions that are destroyed by such disorder. Moreover, we have investigated in detail the critical behavior of the one-dimensional GCP with random-field disorder by means of large-scale Monte Carlo simulations. We have found that the scaling is of activated type comparable to that of the infinite-randomness critical point in the disordered contact process, but with different values of the exponents.  

The main difference between the effects of random-field disorder in the GCP and in equilibrium systems such as random-field Ising model, is the absorbing nature of the inactive states $I_1$ and $I_2$ in the former. The interior of a uniform domain in an equilibrium system (e.g., a spin-up or spin-down domain in the Ising model) can give rise to a new domain of a different state, due to thermal fluctuations. This splits the original domain. Thus, the growth of the typical domain size is limited to its Imry-Ma equilibrium size, resulting in the destruction of the equilibrium transition in sufficiently low dimensions. In contrast, no new domains (nor active sites) can ever, spontaneously, appear in the interior of an $I_1$ or $I_2$ domain. We thus expect that our results are qualitatively valid for all nonequilibrium phase transitions with random-field disorder that locally breaks the symmetry between two \emph{absorbing} states. Actually, Pigolotti and Cencini \cite{PigolottiCencini10} have observed spontaneous symmetry breaking using a model of two competing biological species in a two-dimensional landscape with local habitat preference. The response of other nonequilibrium transitions may be different. For example our theory does not apply to transitions with random fields that break the symmetry between two active states. Furthermore, destabilizing the absorbing character of an inactive state by spontaneous fluctuations, even with small rates, results in the destruction of the phase transition \cite{BorileMaritanMunoz}.

The dynamics in the inactive phase of the GCP with random-field disorder is ultraslow. In one dimension, it is controlled by the Sinai walk of domain walls between the two inactive states. As a result, the densities of domain walls and active sites decay logarithmically with time. The dynamics in two dimensions can be  mapped to that of the well-studied low-temperature random-field Ising model in the regime before the Imry-Ma limit for the domain size is reached. In this regime, the domain wall density decays logarithmically with time. Because an Imry-Ma limit is absent in our system (due to the absorbing nature of the inactive states), this logarithmic time decay of the densities of domain walls and active sites continues for infinite time. Let us also mention the well-studied voter model. In this model each voter (site) can have one of two opinions ($I_1$, $I_2$), and two neighboring voters can convince one another of their own opinion with equal chances. Here, random-field disorder can be introduced in terms of local preference of one opinion over the other. Analogous to the GCP, the dynamics in the random-field one-dimensional voter model is solely controlled by the Sinai walk of domain walls. We therefore expect its dynamics to be, asymptotically, similar to that of the inactive phase of the one-dimensional GCP with random-field disorder.     

We also note that the survival of a nonequilibrium continuous phase transitions in the presence of random-field disorder, implies the survival of the corresponding nonequilibrium first-order phase transition between the two \emph{absorbing} states. (This transition can be tuned through a global preference of one of the two absorbing states.) In contrast, Mart\'{\i}n \emph{et al}. \cite{MartnBonachelaMunoz} have illustrated that nonequilibrium first-order phase transitions between fluctuating and absorbing states are destroyed by quenched disorder, in agreement with the Imry-Ma criterion. 

In the higher-dimensional ($d>2$) GCP, the mapping of domain wall hopping rates onto the random-field Ising model at low-temperatures still holds, but only qualitatively \footnote{In dimensions $d>2$ the number of different configurations of nearest-neighbors in states $I_1$ vs $I_2$ increases. As a result the number of nontrivial hopping rates ratios is more than one and increases with the dimensionality of the system. In the Ising model there is only one parameter to fine tune map these ratios, specifically, the ratio $J/T$.}. In addition, the interior of a uniform absorbing state domain, is still free of any spontaneous fluctuations. Furthermore, the Imry-Ma argument predicts weaker effects of random fields on equilibrium transitions in higher dimensions. All of the above suggests that domain formation will not be able to destroy the absorbing state transition in higher dimensions. However, other unrelated mechanisms may destroy the transition. For example, to the best of our knowledge, not even the clean GCP in dimensions $d>2$ has been studied in detail. Its transition could be destroyed in analogy with the related voter model that never reaches an absorbing state where one opinion dominates, for $d>2$ \cite{Ben-NaimFrachebourgKrapivsky1996}. 

While straightforward experimental realizations of absorbing state transitions were lacking for a long time \cite{Hinrichsen00b}, appealing examples were recently found in driven suspensions \cite{CCGP08,FFGP11}, turbulent liquid crystals \cite{TKCS07}, and superconducting
vortices \cite{OkumaTsugawaMotohashi11}. Moreover, the nonequilibrium nature of biological systems suggests them as potential candidates for observing nonequilibrium transitions. For example experiments in colony biofilms \cite{KXNF11} are accurately represented by a model of two competing strains of bacteria \cite{KorolevNelson11} reveling a transition in the GV universality class (the same class as the clean two-dimensional GCP). 

\section*{Acknowledgements}
This work has been supported in part by the NSF under Grant No. DMR-1205803.

\appendix
\section{DOMAIN WALL HOPPING RATES}
\label{sec:appendixA}
In this Appendix, we map the domain wall hopping rates of the random-field Ising chain in the low-temperature regime onto those of the GCP with random-field disorder in the inactive phase, in one and two space dimensions. We first define state variables $s_i$ for the GCP in analogy to the Ising variables, such that $s_i=-1$ and $s_i=+1$ correspond to site $i$ being in the inactive state $I_1$ and $I_2$, respectively. Also, we denote the decay rates toward the states $I_1$ and $I_2$ at any site $i$ as $\mu_{i}^{(-1)}$ and $\mu_{i}^{(+1)}$, respectively.
\begin{figure}
\includegraphics[width=8.7cm]{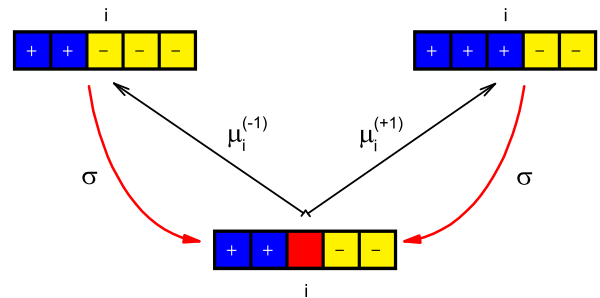}
\caption{Schematics of the dynamics of a $+-$ domain wall in one space dimension. Red, yellow and blue (midtone, light, and dark gray) squares represent a site in the active state $A$, and inactive states $I_1$ and $I_2$ respectively.}
\label{fig:A1}
\end{figure}
Since we are considering only the absorbing phase of the GCP, we can chose the activation rate $\lambda$ to be much smaller than any other rate in the system such that the activation process (\ref{eq:rate_lambda}) can be ignored.

First we consider the mapping in one space dimension. Figure \ref{fig:A1} shows a schematic of the ($+-$) domain wall dynamics. As shown in the figure, the hopping of the domain wall across site $i$ from left to right and from right to left is based on two consecutive processes, an activation of site $i$ through the boundary activation process (\ref{eq:rate_sigma}) with probability rate $\sigma$, followed by a decay toward an inactive states ($I_1$or $I_2$)  with a total decay rate $\mu_{i}^{(-1)}+\mu_{i}^{(+1)}$. The total effective probability rate $w$ of the two consecutive processes behaves as the inverse of their typical total time $\tau_1+\tau_2$ where $\tau_1=1/\sigma$ and $\tau_2=1/(\mu_{i}^{(-1)}+\mu_{i}^{(+1)})$.
i.e., $w=1/(\tau_1+\tau_2)$. The outcome of this combined process is the  hopping of the domain wall from the right (left) to the left (right) of site $i$ with probability of $\mu_{i}^{(-1)}\tau_2$ ($\mu_{i}^{(+1)}\tau_2$) provided that the active site $i$ decays to an inactive state that is different than the initial one. However, with probability of $\mu_{i}^{(+1)}\tau_2$ ($\mu_{i}^{(-1)}\tau_2$)  the decay process leaves site $i$ in the same initial inactive state, i.e., the domain wall does not move. The hopping rates $w(\leftarrow)$ and $w(\rightarrow)$ to the left (right) can be found by multiplying $w$ with the probability that the active site $i$ ends up in a different inactive state than the initial one. Doing so we get
\begin{equation}
w(\leftarrow)=\dfrac{\sigma\mu_{i}^{(-1)}}{\sigma+\mu_{i}^{(-1)}+\mu_{i}^{(+1)}},
\label{eq:w_left}
\end{equation}  
\begin{equation}
w(\rightarrow)=\dfrac{\sigma\mu_{i}^{(+1)}}{\sigma+\mu_{i}^{(+1)}+\mu_{i}^{(-1)}}.
\label{eq:w_right}
\end{equation}
In general we can write        
\begin{equation}
w(\leftarrow)=\dfrac{\sigma\mu_{i}^{(-s_i)}}{\sigma+\mu_{i}^{(-s_i)}+\mu_{i}^{(s_i)}},
\label{eq:w_left_gen}
\end{equation}  
\begin{equation}
w(\rightarrow)=\dfrac{\sigma\mu_{i}^{(s_{i})}}{\sigma+\mu_{i}^{(-s_{i})}+\mu_{i}^{(s_{i})}},
\label{eq:w_right_gen}
\end{equation}
where $s_i$ is the state of site $i$ when it is to the left of the domain wall.
The ratio of the hopping rates is
\begin{equation}
\dfrac{w(\rightarrow)}{w(\leftarrow)}=\dfrac{\mu_{i}^{(s_{i})}}{\mu_{i}^{(-s_i)}}.
\label{eq:w_ratio}
\end{equation}  
Using the variable $\alpha_i=\ln(\mu_{i}^{(+1)}/\mu_{i}^{(-1)})$, we can write $s_i\alpha_i=\ln(\mu_{i}^{(s_i)}/\mu_{i}^{(-s_i)})$. This lead to
\begin{equation}
\dfrac{w(\rightarrow)}{w(\leftarrow)}=\exp(s_i\alpha_{i}).
\label{eq:w_ratio_alpha}
\end{equation}

Now, we turn to the case of the random-field Ising model defined by the Hamiltonian
\begin{equation}
H=-J\sum_{<i,j>}s_is_j-\sum_ih_is_i,
\label{eq:RFIM_H}
\end{equation}
where $J>0$ and $h_i$ is a random variable drawn from a symmetric distribution  such that $<h_i>=0$. The transition rates ratio can be found from the detailed balance equation as
\begin{equation}
\dfrac{w(\rightarrow)}{w(\leftarrow)}=\exp(-\Delta E/T)=\exp(2s_ih_i/T),
\label{eq:w_RFIM_right/left}
\end{equation}
where $\Delta E$ is the change in the system energy as the spin at site $i$ flips from $-s_i$ to $s_i$.
From Eq. (\ref{eq:w_ratio_alpha}) and Eq. (\ref{eq:w_RFIM_right/left}) the two
systems have equal  hopping rate ratios if
\begin{equation}
\alpha_i=2h_i/T.
\label{eq:mapping_alpha_h}
\end{equation}
For the binary distribution (\ref{eq:distrib}) and (\ref{eq:alpha_distrib}), this implies
\begin{equation}
\ln(\mu_h/\mu_l)=\alpha_0=2h_0/T,
\label{eq:mapping_final_0}
\end{equation}
where $h_i$ is drawn from a symmetric binary distribution with possible values of $\pm h_0$. 

We now turn to two dimensions. In contrast to the one-dimensional case where the domain wall size is fixed (it always consists of a single $+-$ bond); domain walls in two space dimensions may change size length as the hopping of a domain wall segment might result in the creation of new segments or the annihilation of existing ones
\begin{figure}
\includegraphics[width=8.7cm]{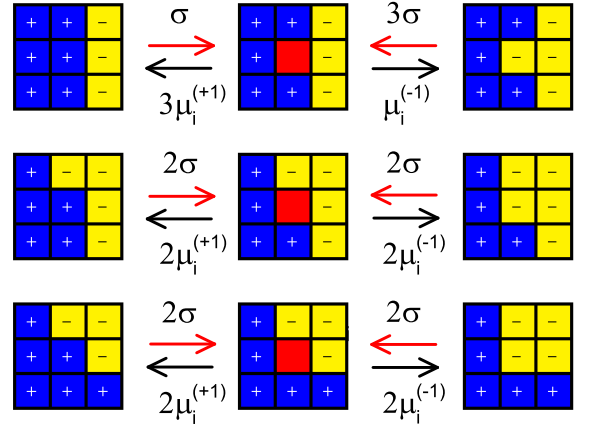}
\caption{Schematics of the dynamics of a $+-$ domain wall in two space dimensions. Red, yellow, and blue (midtone, light, and dark gray) squares represent a site in an active state $A$ and inactive states $I_1$ and $I_2$ respectively.}
\label{fig:A2}
\end{figure}
(Fig. \ref{fig:A2}). Therefore, the domain wall geometry must be taken into account. We consider the domain wall motion due to a single site changing from $+1$ to $-1$ or from $-1$ to $+1$, as sketched in Fig. \ref{fig:A2}. As in the one-dimensional case, the hopping consists of two consecutive processes. First the inactive site $i$ in the inactive state $s_i$ must be activated with probability rate of $n_{\rm{dif}}\sigma$ followed by a decay toward an inactive states $-s_i$ or $s_i$  with a total decay rate $n_{\rm{dif}}\mu_{i}^{(-s_i)}+(4-n_{\rm{dif}})\mu_{i}^{(s_i)}$. Here, $n_{\rm{dif}}$ is the number of inactive neighbors in a different state than $s_i$, i.e., in state $-s_i$. (In order to suppress any activation of one of the neighbors of site $i$ before the decay of site $i$ to an inactive state, we work in the limit $\mu_i^{\pm s_i}/\sigma\gg1$.) The effective probability rate $w$ of the two consecutive processes is
\begin{equation}
w=\dfrac{n_{\rm{dif}}\sigma[n_{\rm{dif}}\mu_{i}^{(-s_i)}+(4-n_{\rm{dif}})\mu_{i}^{(s_i)}]}{n_{\rm{dif}}\sigma+n_{\rm{dif}}\mu_{i}^{(-s_i)}+(4-n_{\rm{dif}})\mu_{i}^{(s_i)}}.
\label{eq:w}
\end{equation}
The probability that site $i$ will end up in a different inactive state than the initial one at the end of this process is $n_{\rm{dif}}\mu_{i}^{(-s_i)}/(n_{\rm{dif}}\mu_{i}^{(-s_i)}+(4-n_{\rm{dif}})\mu_{i}^{(s_i)})$. As a result, the hopping rate $w_{s_i\rightarrow-s_i}$ from state $s_i$ to $-s_i$ is 
\begin{equation}
w_{s_i\rightarrow-s_i}=\dfrac{n_{\rm{dif}}^2\sigma\mu_{i}^{(-s_i)}}{n_{\rm{dif}}\sigma+n_{\rm{dif}}\mu_{i}^{(-s_i)}+(4-n_{\rm{dif}})\mu_{i}^{(s_i)}}
\label{eq:w+-}~~.
\end{equation}
The hopping rate of site $i$ back to its initial state $w_{-s_i\rightarrow s_i}$ can simply be found by interchanging $s_i$ with $-s_i$ and $n_{\rm{dif}}$ with $4-n_{\rm{dif}}$ in Eq. (\ref{eq:w+-}):
\begin{equation}
w_{-s_i\rightarrow s_i}=\dfrac{(4-n_{\rm{dif}})^2\sigma\mu_{i}^{(s_i)}}{(4-n_{\rm{dif}})\sigma+(4-n_{\rm{dif}})\mu_{i}^{(s_i)}+n_{\rm{dif}}\mu_{i}^{(-s_i)}}
\label{eq:w-+}~~.
\end{equation}
The ratio between $s_i\rightarrow-s_i$ and $-s_i\rightarrow s_i$ hopping rates is
\begin{eqnarray}
\dfrac{w_{s_i\rightarrow-s_i}}{w_{-s_i\rightarrow s_i}}=\left(\dfrac{n_{\rm{dif}}}{4-n_{\rm{dif}}}\right)^2
\left(\dfrac{\mu_{i}^{(-s_i)}}{\mu_{i}^{(s_i)}}\right)\hspace{3.0cm}&\nonumber\\
\times\left[\dfrac{(4-n_{\rm{dif}})\sigma+(4-n_{\rm{dif}})\mu_{i}^{(s_i)}+n_{\rm{dif}}\mu_{i}^{(-s_i)}}{n_{\rm{dif}}\sigma+n_{\rm{dif}}\mu_{i}^{(-s_i)}+(4-n_{\rm{dif}})\mu_{i}^{(s_i)}}\right].&\nonumber\\
\label{eq:w+-to-+}
\end{eqnarray}
In the limit $\mu_i^{\pm s_i}/\sigma\gg1$, the right most factor of Eq. (\ref{eq:w+-to-+}) is equal to unity, the middle factor is similar to the random-field factor in the one-dimensional case [Eq. (\ref{eq:w_ratio})] and the first factor encodes the geometry. Therefore, we can write,
\begin{equation}
\dfrac{w_{s_i\rightarrow-s_i}}{w_{-s_i\rightarrow s_i}}=\left(\dfrac{n_{\rm{dif}}}{4-n_{\rm{dif}}}\right)^2
\exp(s_i\alpha_{i})
\label{eq:w+-to-+_s}.
\end{equation}
Considering the possible values of $n_{\rm{dif}}$ for a site at a domain wall, we get: 
\begin{equation}
\dfrac{w_{s_i\rightarrow-s_i}}{w_{-s_i\rightarrow s_i}}= \left\{
\begin{array}{cc} 1/9\exp(s_i\alpha_{i}) & (n_{\rm{dif}}=1) \\
                  \exp(s_i\alpha_{i})~~~~~    & (n_{\rm{dif}}=2) \\
                  9\exp(s_i\alpha_{i})~~~   & (n_{\rm{dif}}=3)
\end{array}.
\right.
\label{eq:w+-to-+_123}
\end{equation}

In the case of two-dimensional random-field Ising model, the transition rates ratio can be found from the detailed balance equation as
\begin{equation}
\dfrac{w_{s_i\rightarrow-s_i}}{w_{-s_i\rightarrow s_i}}=\exp[4(n_{\rm{dif}}-2)J/T+2s_ih_i/T]
\label{eq:w+-to-+RFIM},
\end{equation}
substituting for the possible values of $n_{\rm{dif}}$ we get: 
\begin{equation}
\dfrac{w_{s_i\rightarrow-s_i}}{w_{-s_i\rightarrow s_i}}= \left\{
\begin{array}{cc} \exp(-4J/T+s_ih_{i}/T)&(n_{\rm{dif}}=1) \\
                  \exp(s_ih_{i}/T)~~~~~~~~~~~~~& (n_{\rm{dif}}=2) \\
                  \exp(4J/T+s_ih_{i}/T)~~& (n_{\rm{dif}}=3)
\end{array}.
\right.
\label{eq:w+-to-+RFIM_123}
\end{equation}
The comparison between Eq. (\ref{eq:w+-to-+_123}) and Eq. (\ref{eq:w+-to-+RFIM_123}), suggests the same mapping of the random-field term as in the one-dimensional case,
\begin{equation}
h_i/T=\alpha_i/2,
\label{eq:mapping_alpha_h_2d}
\end{equation}
while the ratio $J/T$ is constant,
\begin{equation}
J/T=\ln(3)/2.
\label{eq:mapping_J_2d}
\end{equation}
\section{DOMAIN WALL DYNAMICS IN THE RFIM}
\label{sec:appendixB}
Here we consider the random-field Ising model defined by the Hamiltonian,
\begin{equation}
H=-J\sum_{<i,j>}s_is_j-\sum_ih_is_i,
\label{eq:RFIM_H_2d}
\end{equation}
where $\langle h_i\rangle=0$ and $\langle h_ih_j\rangle=h^2\delta_{i,j}$ in the limit $h \ll J$.
(The results in this appendix have been derived in Refs. \cite{Villain84,Grinstein84,Grinstein83}, we summarize them for the convenience of the reader.)

\subsection{Interface roughening in the RFIM}
\label{sec:roughening}
\begin{figure}
\includegraphics[width=8.7cm]{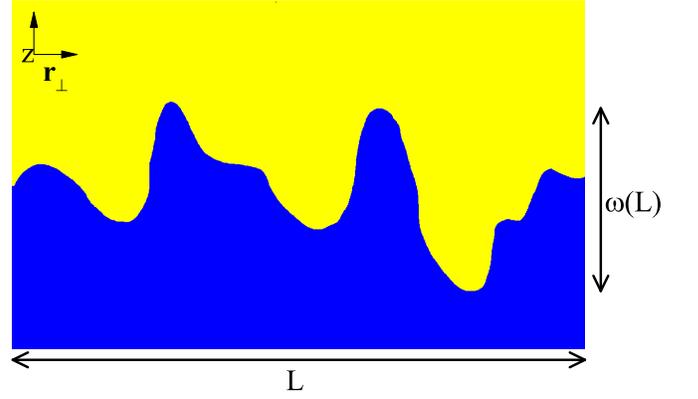}
\caption{Interface separating domains of spin-up [blue (dark gray)] and spin-down [yellow (light gray)] with interface profile $z(\mathbf{r}_\bot)$.}
\label{fig:B1}
\end{figure}

In the absence of disorder ($h=0$) the interface between spin-up and spin-down domains will tend to be flat in order to minimize the surface energy $E_J$. However, random-field disorder prefers an interface profile that follows the random-field fluctuations in order to minimize the field energy $E_h$. Let $z(\mathbf{r}_\bot)$ be the interface profile function (Fig. \ref{fig:B1}). The increase of the surface energy compared to its flat interface value can be estimated as
\begin{equation}
\Delta E_J \sim J\int d^{d-1}\mathbf{r}_\bot\left[1+(\nabla z)^2 \right]^{1/2}-J\int d^{d-1}\mathbf{r}_\bot.
\label{eq:Delta-E_J}
\end{equation}
If $z(\mathbf{r}_\bot)$ fluctuates on a scale of $\omega(L)$, where $\omega(L) \ll L$, we can approximate $\left(1+(\nabla z)^2 \right)^{1/2}$ by $1+1/2[\omega(L)/L]^2$ to obtain 
\begin{equation}
\Delta E_J \sim JL^{d-3}\omega^2(L).
\label{eq:Delta-E_J_w}
\end{equation}
The gain in random field due to reshaping the interface to a favorable profile is (based on a central limit theorem argument) proportional to the square root of the interface volume and $h$ such that    
\begin{equation}
\Delta E_h \sim -h\left[L^{d-1}\omega(L)\right]^{1/2}.
\label{eq:Delta-E_h_w}
\end{equation}
If we minimize the total energy change $\Delta E=\Delta E_J+\Delta E_h$ with respect to $\omega(L)$, we get
\begin{equation}
\omega_{\rm{min}}\sim \left(h/J\right)^{2/3}L^{(5-d)/3},
\label{eq:omega_min}
\end{equation}

which corresponds to energy gain of
\begin{equation}
\Delta E_{\rm{min}} \sim J\left(h/J\right)^{4/3}L^{(d+1)/3}.
\label{eq:Delta-E_min}
\end{equation}
Based on Eq. (\ref{eq:omega_min}) the interface width ($\omega_{\rm{min}}$) is bounded (smooth) for $d>5$ and infinitely increasing for $d<5$, where 
\begin{equation}
\lim_{L\to\infty} \omega_{\rm{min}}= \left\{
\begin{array}{cc} 0 & (d>5) \\
                  \infty& (d<5)
\end{array},
\right. 
\label{eq:omega_min_limits}
\end{equation}
However, the ratio 
\begin{equation}
\dfrac{\omega_{\rm{min}}}{L}\sim \left(h/J\right)^{2/3}L^{(2-d)/3}
\label{eq:omega_min_byL}
\end{equation} 
is bounded for $d>2$, where
\begin{equation}
\lim_{L\to\infty} \dfrac{\omega_{\rm{min}}}{L}= \left\{
\begin{array}{cc} 0 & (d>2) \\
                  \infty& (d<2)
\end{array}.
\right.
\label{eq:omega_min_by_L}
\end{equation}
Accordingly, the interface is rough on scale of $w(L) \ll L$ for $2<d<5$. 

\subsection{Asymptotic Interface Dynamics}
Consider a spherical $d$-dimensional spin-up (spin-down) domain of radius $R$ embedded in a much larger spin-down (spin-up) domain. Also, consider that the interface profile minimizes the random-field energy locally (the interface is in a favorable position w.r.t. the random field). According to the above results, the interface is rough on a scale $w \ll R$ for $2<d<5$. The embedded domain wishes to reduce the surface energy by shrinking but the random-field creates an energy barrier against the interface motion.

In order to estimate the energy barrier height, we assume that the radius of the embedded domain shrinks from $R$ to $R-\Delta r$. As a result, the surface energy will decrease as
\begin{equation}
\Delta E_J \sim -JR^{d-2}\Delta r. 
\label{eq:dE_J}
\end{equation}
As the interface moves it covers a volume proportional to $R^{d-1}\Delta r$. The typical value of the random-field energy in an unfavorable configuration is
\begin{equation}
\Delta E_{h} \sim h\left(R^{d-1}\Delta r\right)^{1/2}.
\label{eq:dE_h}
\end{equation}
The total energy change is then 
\begin{equation}
\Delta E \sim -JR^{d-2}\Delta r+h\left(R^{d-1}\Delta r\right)^{1/2},
\label{eq:dE}
\end{equation}
where proportionality factors are suppressed. As $\Delta r$ starts to increase from  zero, the random-field term $\Delta E_h$ initially dominates over the surface term $\Delta E_J$ in Eq. (\ref{eq:dE}). As $\Delta r$ continues to increase the surface term will win eventually, and the interface reaches a new favorable position w.r.t. the random field. The typical height of the energy barrier can be found by maximizing $\Delta E$ given in Eq. (\ref{eq:dE}). This leads to a barrier height of    
\begin{equation}
\Delta E_{\rm{max}} \sim h^2R/(4J)
\label{eq:dE_max},
\end{equation}
with a typical width of 
\begin{equation}
\Delta r_{\rm{max}} \sim h^2R^{3-d}/(4J^2)
\label{eq:dr_max}.
\end{equation}
The time taken to overcome an energy barrier of height $\Delta E$ at temperature $T$ depends exponentially on the ratio $\Delta E/T$, i.e.,
\begin{equation}
t=t_0\exp(\Delta E/T)
\label{eq:t_Asymptotic},
\end{equation}
where $t_0$ is a microscopic time scale.
This means that at time $t$, energy barriers lower than $T\ln(t/t_0)$
have been overcome, while energy barriers higher than $T\ln(t/t_0)$ have not yet been overcome. Therefore, the typical domain radius $R$ [based on Eq. (\ref{eq:dE_max})] at time $t$ is  
\begin{equation}
R \sim (JT/h^2)\ln(t/t_0)
\label{eq:R_t_Asymptotic}.
\end{equation}
Note that smaller domains have been eliminated by shrinking; (when a domain starts shrinking, it collapses because the smaller the radius the lower the barrier).

\subsection{Pre-asymptotic Interface Dynamics}
The previous results are in the asymptotic regime $R\gg1$, where all energies are much greater than the microscopic scales $J$ and $h$. In this case, treating $\Delta E$ as continuous is justified. However, the change in the interface energy $\Delta E_J$ cannot be less than $J$. This means that Eq. (\ref{eq:dE_max}), which governs the dependence of the barrier height $\Delta E_{\rm{max}}$ on $R$, break down as $R$ decreases below the crossover value $R_x\sim J^2/h^2$.

\begin{figure}
\includegraphics[width=8.7cm]{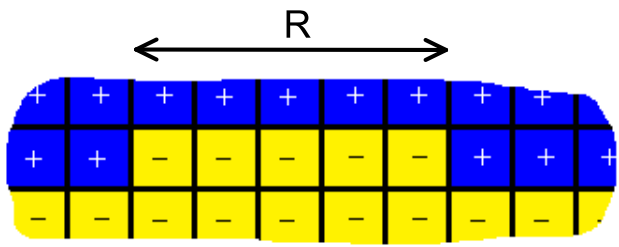}
\caption{Interface separating domains of spin-up [blue (dark gray)] and spin-down [yellow (light gray)] with a double kink of spin-down  on top of otherwise flat interface.}
\label{fig:B2}
\end{figure}

In this regime($R\ll R_x$), microscopic considerations must be taken into account. First, we consider this regime in two dimensions. Start with a domain wall that is flat except for a double kink as shown in Fig. \ref{fig:B2}.
Only spins right next to the kink can flip without increasing the interface length . A spin flip that increases the interface length costs energy of order of $J \gg h$. Therefore, it is unlikely to happen. Instead, sides of the double kink can move left and right, with probabilities that depend only on the random-field values, until they meet and cancel each other. Before the sides of the double kink meet there is no gain in the interface energy $\Delta E_J$ . However, there is energy cost (barrier) of $\Delta E_h \sim hR^{1/2}$ where $R$ is the distance between the kinks. The characteristic decay time is $t=t_0\exp(\Delta E_h/T)$, which leads to
\begin{equation}
R \sim (T^2/h^2)\ln^2(t/t_0)
\label{eq:R_t_pre_asymptotic}.
\end{equation}
In higher dimensions ($d>2$), we consider an island of size $R^{d-1}$ on top of otherwise flat interface. In this case the elimination of such an island can be done by eliminating one-dimensional rows in any of the ($d-1$) directions at a time. Each row elimination involves a barrier of $hR^{1/2}$ and result in an energy gain of $\sim J$. Therefore, two-dimensional results applies for all $d \ge 2$. In summary, the typical domain radius behaves as 
\begin{equation}
R\sim \left\{
\begin{array}{cc}(T^2/h^2)\ln^2(t/t_0) & (t<t_x) \\\\
                  (JT/h^2)\ln(t/t_0)& (t>t_x)
\end{array},
\right. 
\label{eq:R_t_general}
\end{equation}
where
\begin{equation}
\ln(t_x/t_0) \sim J/T
\label{eq:t_x_}.
\end{equation}

\bibliographystyle{apsrev4-1}
\bibliography{../../nonequilibrium}

\begin{thebibliography}{59}%
\makeatletter
\providecommand \@ifxundefined [1]{%
 \@ifx{#1\undefined}
}%
\providecommand \@ifnum [1]{%
 \ifnum #1\expandafter \@firstoftwo
 \else \expandafter \@secondoftwo
 \fi
}%
\providecommand \@ifx [1]{%
 \ifx #1\expandafter \@firstoftwo
 \else \expandafter \@secondoftwo
 \fi
}%
\providecommand \natexlab [1]{#1}%
\providecommand \enquote  [1]{``#1''}%
\providecommand \bibnamefont  [1]{#1}%
\providecommand \bibfnamefont [1]{#1}%
\providecommand \citenamefont [1]{#1}%
\providecommand \href@noop [0]{\@secondoftwo}%
\providecommand \href [0]{\begingroup \@sanitize@url \@href}%
\providecommand \@href[1]{\@@startlink{#1}\@@href}%
\providecommand \@@href[1]{\endgroup#1\@@endlink}%
\providecommand \@sanitize@url [0]{\catcode `\\12\catcode `\$12\catcode
  `\&12\catcode `\#12\catcode `\^12\catcode `\_12\catcode `\%12\relax}%
\providecommand \@@startlink[1]{}%
\providecommand \@@endlink[0]{}%
\providecommand \url  [0]{\begingroup\@sanitize@url \@url }%
\providecommand \@url [1]{\endgroup\@href {#1}{\urlprefix }}%
\providecommand \urlprefix  [0]{URL }%
\providecommand \Eprint [0]{\href }%
\providecommand \doibase [0]{http://dx.doi.org/}%
\providecommand \selectlanguage [0]{\@gobble}%
\providecommand \bibinfo  [0]{\@secondoftwo}%
\providecommand \bibfield  [0]{\@secondoftwo}%
\providecommand \translation [1]{[#1]}%
\providecommand \BibitemOpen [0]{}%
\providecommand \bibitemStop [0]{}%
\providecommand \bibitemNoStop [0]{.\EOS\space}%
\providecommand \EOS [0]{\spacefactor3000\relax}%
\providecommand \BibitemShut  [1]{\csname bibitem#1\endcsname}%
\let\auto@bib@innerbib\@empty
\bibitem [{\citenamefont {Harris}\ and\ \citenamefont
  {Lubensky}(1974)}]{HarrisLubensky74}%
  \BibitemOpen
  \bibfield  {author} {\bibinfo {author} {\bibfnamefont {A.~B.}\ \bibnamefont
  {Harris}}\ and\ \bibinfo {author} {\bibfnamefont {T.~C.}\ \bibnamefont
  {Lubensky}},\ }\href@noop {} {\bibfield  {journal} {\bibinfo  {journal}
  {Phys. Rev. Lett.}\ }\textbf {\bibinfo {volume} {33}},\ \bibinfo {pages}
  {1540} (\bibinfo {year} {1974})}\BibitemShut {NoStop}%
\bibitem [{\citenamefont {Grinstein}\ and\ \citenamefont
  {Luther}(1976)}]{GrinsteinLuther76}%
  \BibitemOpen
  \bibfield  {author} {\bibinfo {author} {\bibfnamefont {G.}~\bibnamefont
  {Grinstein}}\ and\ \bibinfo {author} {\bibfnamefont {A.}~\bibnamefont
  {Luther}},\ }\href@noop {} {\bibfield  {journal} {\bibinfo  {journal} {Phys.
  Rev. B}\ }\textbf {\bibinfo {volume} {13}},\ \bibinfo {pages} {1329}
  (\bibinfo {year} {1976})}\BibitemShut {NoStop}%
\bibitem [{\citenamefont {Vojta}(2003{\natexlab{a}})}]{Vojta03a}%
  \BibitemOpen
  \bibfield  {author} {\bibinfo {author} {\bibfnamefont {T.}~\bibnamefont
  {Vojta}},\ }\href@noop {} {\bibfield  {journal} {\bibinfo  {journal} {Phys.
  Rev. Lett.}\ }\textbf {\bibinfo {volume} {90}},\ \bibinfo {pages} {107202}
  (\bibinfo {year} {2003}{\natexlab{a}})}\BibitemShut {NoStop}%
\bibitem [{\citenamefont {Vojta}(2003{\natexlab{b}})}]{Vojta03b}%
  \BibitemOpen
  \bibfield  {author} {\bibinfo {author} {\bibfnamefont {T.}~\bibnamefont
  {Vojta}},\ }\href@noop {} {\bibfield  {journal} {\bibinfo  {journal} {J.
  Phys. A}\ }\textbf {\bibinfo {volume} {36}},\ \bibinfo {pages} {10921}
  (\bibinfo {year} {2003}{\natexlab{b}})}\BibitemShut {NoStop}%
\bibitem [{\citenamefont {Vojta}(2004)}]{Vojta04}%
  \BibitemOpen
  \bibfield  {author} {\bibinfo {author} {\bibfnamefont {T.}~\bibnamefont
  {Vojta}},\ }\href@noop {} {\bibfield  {journal} {\bibinfo  {journal} {Phys.
  Rev. E}\ }\textbf {\bibinfo {volume} {70}},\ \bibinfo {pages} {026108}
  (\bibinfo {year} {2004})}\BibitemShut {NoStop}%
\bibitem [{\citenamefont {Imry}\ and\ \citenamefont
  {Wortis}(1979)}]{ImryWortis79}%
  \BibitemOpen
  \bibfield  {author} {\bibinfo {author} {\bibfnamefont {Y.}~\bibnamefont
  {Imry}}\ and\ \bibinfo {author} {\bibfnamefont {M.}~\bibnamefont {Wortis}},\
  }\href {\doibase 10.1103/PhysRevB.19.3580} {\bibfield  {journal} {\bibinfo
  {journal} {Phys. Rev. B}\ }\textbf {\bibinfo {volume} {19}},\ \bibinfo
  {pages} {3580} (\bibinfo {year} {1979})}\BibitemShut {NoStop}%
\bibitem [{\citenamefont {Hui}\ and\ \citenamefont
  {Berker}(1989)}]{HuiBerker89}%
  \BibitemOpen
  \bibfield  {author} {\bibinfo {author} {\bibfnamefont {K.}~\bibnamefont
  {Hui}}\ and\ \bibinfo {author} {\bibfnamefont {A.~N.}\ \bibnamefont
  {Berker}},\ }\href {\doibase 10.1103/PhysRevLett.62.2507} {\bibfield
  {journal} {\bibinfo  {journal} {Phys. Rev. Lett.}\ }\textbf {\bibinfo
  {volume} {62}},\ \bibinfo {pages} {2507} (\bibinfo {year}
  {1989})}\BibitemShut {NoStop}%
\bibitem [{\citenamefont {Aizenman}\ and\ \citenamefont
  {Wehr}(1989)}]{AizenmanWehr89}%
  \BibitemOpen
  \bibfield  {author} {\bibinfo {author} {\bibfnamefont {M.}~\bibnamefont
  {Aizenman}}\ and\ \bibinfo {author} {\bibfnamefont {J.}~\bibnamefont
  {Wehr}},\ }\href {\doibase 10.1103/PhysRevLett.62.2503} {\bibfield  {journal}
  {\bibinfo  {journal} {Phys. Rev. Lett.}\ }\textbf {\bibinfo {volume} {62}},\
  \bibinfo {pages} {2503} (\bibinfo {year} {1989})}\BibitemShut {NoStop}%
\bibitem [{\citenamefont {Tabei}\ \emph {et~al.}(2006)\citenamefont {Tabei},
  \citenamefont {Gingras}, \citenamefont {Kao}, \citenamefont {Stasiak},\ and\
  \citenamefont {Fortin}}]{TGKSF06}%
  \BibitemOpen
  \bibfield  {author} {\bibinfo {author} {\bibfnamefont {S.~M.~A.}\
  \bibnamefont {Tabei}}, \bibinfo {author} {\bibfnamefont {M.~J.~P.}\
  \bibnamefont {Gingras}}, \bibinfo {author} {\bibfnamefont {Y.-J.}\
  \bibnamefont {Kao}}, \bibinfo {author} {\bibfnamefont {P.}~\bibnamefont
  {Stasiak}}, \ and\ \bibinfo {author} {\bibfnamefont {J.-Y.}\ \bibnamefont
  {Fortin}},\ }\href@noop {} {\bibfield  {journal} {\bibinfo  {journal} {Phys.
  Rev. Lett.}\ }\textbf {\bibinfo {volume} {97}},\ \bibinfo {eid} {237203}
  (\bibinfo {year} {2006})}\BibitemShut {NoStop}%
\bibitem [{\citenamefont {Silevitch}\ \emph {et~al.}(2007)\citenamefont
  {Silevitch}, \citenamefont {Bitko}, \citenamefont {Brooke}, \citenamefont
  {Ghosh}, \citenamefont {Aeppli},\ and\ \citenamefont {Rosenbaum}}]{SBBGAR07}%
  \BibitemOpen
  \bibfield  {author} {\bibinfo {author} {\bibfnamefont {D.~M.}\ \bibnamefont
  {Silevitch}}, \bibinfo {author} {\bibfnamefont {D.}~\bibnamefont {Bitko}},
  \bibinfo {author} {\bibfnamefont {J.}~\bibnamefont {Brooke}}, \bibinfo
  {author} {\bibfnamefont {S.}~\bibnamefont {Ghosh}}, \bibinfo {author}
  {\bibfnamefont {G.}~\bibnamefont {Aeppli}}, \ and\ \bibinfo {author}
  {\bibfnamefont {T.~F.}\ \bibnamefont {Rosenbaum}},\ }\href@noop {} {\bibfield
   {journal} {\bibinfo  {journal} {Nature}\ }\textbf {\bibinfo {volume}
  {448}},\ \bibinfo {pages} {567} (\bibinfo {year} {2007})}\BibitemShut
  {NoStop}%
\bibitem [{\citenamefont {Schechter}(2008)}]{Schechter08}%
  \BibitemOpen
  \bibfield  {author} {\bibinfo {author} {\bibfnamefont {M.}~\bibnamefont
  {Schechter}},\ }\href@noop {} {\bibfield  {journal} {\bibinfo  {journal}
  {Phys. Rev. B}\ }\textbf {\bibinfo {volume} {77}},\ \bibinfo {eid} {020401}
  (\bibinfo {year} {2008})}\BibitemShut {NoStop}%
\bibitem [{\citenamefont {Maritan}\ \emph {et~al.}(1994)\citenamefont
  {Maritan}, \citenamefont {Cieplak}, \citenamefont {Bellini},\ and\
  \citenamefont {Banavar}}]{MCBB94}%
  \BibitemOpen
  \bibfield  {author} {\bibinfo {author} {\bibfnamefont {A.}~\bibnamefont
  {Maritan}}, \bibinfo {author} {\bibfnamefont {M.}~\bibnamefont {Cieplak}},
  \bibinfo {author} {\bibfnamefont {T.}~\bibnamefont {Bellini}}, \ and\
  \bibinfo {author} {\bibfnamefont {J.~R.}\ \bibnamefont {Banavar}},\ }\href
  {\doibase 10.1103/PhysRevLett.72.4113} {\bibfield  {journal} {\bibinfo
  {journal} {Phys. Rev. Lett.}\ }\textbf {\bibinfo {volume} {72}},\ \bibinfo
  {pages} {4113} (\bibinfo {year} {1994})}\BibitemShut {NoStop}%
\bibitem [{\citenamefont {Carlson}\ \emph {et~al.}(2006)\citenamefont
  {Carlson}, \citenamefont {Dahmen}, \citenamefont {Fradkin},\ and\
  \citenamefont {Kivelson}}]{CDFK06}%
  \BibitemOpen
  \bibfield  {author} {\bibinfo {author} {\bibfnamefont {E.~W.}\ \bibnamefont
  {Carlson}}, \bibinfo {author} {\bibfnamefont {K.~A.}\ \bibnamefont {Dahmen}},
  \bibinfo {author} {\bibfnamefont {E.}~\bibnamefont {Fradkin}}, \ and\
  \bibinfo {author} {\bibfnamefont {S.~A.}\ \bibnamefont {Kivelson}},\ }\href
  {\doibase 10.1103/PhysRevLett.96.097003} {\bibfield  {journal} {\bibinfo
  {journal} {Phys. Rev. Lett.}\ }\textbf {\bibinfo {volume} {96}},\ \bibinfo
  {pages} {097003} (\bibinfo {year} {2006})}\BibitemShut {NoStop}%
\bibitem [{\citenamefont {Imry}\ and\ \citenamefont {Ma}(1975)}]{ImryMa75}%
  \BibitemOpen
  \bibfield  {author} {\bibinfo {author} {\bibfnamefont {Y.}~\bibnamefont
  {Imry}}\ and\ \bibinfo {author} {\bibfnamefont {S.-k.}\ \bibnamefont {Ma}},\
  }\href {\doibase 10.1103/PhysRevLett.35.1399} {\bibfield  {journal} {\bibinfo
   {journal} {Phys. Rev. Lett.}\ }\textbf {\bibinfo {volume} {35}},\ \bibinfo
  {pages} {1399} (\bibinfo {year} {1975})}\BibitemShut {NoStop}%
\bibitem [{Note1()}]{Note1}%
  \BibitemOpen
  \bibinfo {note} {This holds for discrete symmetry. For continuous symmetry
  the surface energy behaves as $L^{d-2}$ resulting in a marginal dimension of
  4}\BibitemShut {NoStop}%
\bibitem [{\citenamefont {Schmittmann}\ and\ \citenamefont
  {Zia}(1995)}]{SchmittmannZia95}%
  \BibitemOpen
  \bibfield  {author} {\bibinfo {author} {\bibfnamefont {B.}~\bibnamefont
  {Schmittmann}}\ and\ \bibinfo {author} {\bibfnamefont {R.~K.~P.}\
  \bibnamefont {Zia}},\ }in\ \href@noop {} {\emph {\bibinfo {booktitle} {Phase
  Transitions and Critical Phenomena}}},\ Vol.~\bibinfo {volume} {17},\
  \bibinfo {editor} {edited by\ \bibinfo {editor} {\bibfnamefont
  {C.}~\bibnamefont {Domb}}\ and\ \bibinfo {editor} {\bibfnamefont {J.~L.}\
  \bibnamefont {Lebowitz}}}\ (\bibinfo  {publisher} {Academic},\ \bibinfo
  {address} {New York},\ \bibinfo {year} {1995})\ p.~\bibinfo {pages}
  {1}\BibitemShut {NoStop}%
\bibitem [{\citenamefont {Marro}\ and\ \citenamefont
  {Dickman}(1999)}]{MarroDickman99}%
  \BibitemOpen
  \bibfield  {author} {\bibinfo {author} {\bibfnamefont {J.}~\bibnamefont
  {Marro}}\ and\ \bibinfo {author} {\bibfnamefont {R.}~\bibnamefont
  {Dickman}},\ }\href@noop {} {\emph {\bibinfo {title} {Nonequilibrium Phase
  Transitions in Lattice Models}}}\ (\bibinfo  {publisher} {Cambridge
  University Press},\ \bibinfo {address} {Cambridge},\ \bibinfo {year}
  {1999})\BibitemShut {NoStop}%
\bibitem [{\citenamefont {Hinrichsen}(2000{\natexlab{a}})}]{Hinrichsen00}%
  \BibitemOpen
  \bibfield  {author} {\bibinfo {author} {\bibfnamefont {H.}~\bibnamefont
  {Hinrichsen}},\ }\href {\doibase 10.1080/00018730050198152} {\bibfield
  {journal} {\bibinfo  {journal} {Adv. Phys.}\ }\textbf {\bibinfo {volume}
  {49}},\ \bibinfo {pages} {815} (\bibinfo {year}
  {2000}{\natexlab{a}})}\BibitemShut {NoStop}%
\bibitem [{\citenamefont {Odor}(2004)}]{Odor04}%
  \BibitemOpen
  \bibfield  {author} {\bibinfo {author} {\bibfnamefont {G.}~\bibnamefont
  {Odor}},\ }\href {\doibase 10.1103/RevModPhys.76.663} {\bibfield  {journal}
  {\bibinfo  {journal} {Rev. Mod. Phys.}\ }\textbf {\bibinfo {volume} {76}},\
  \bibinfo {pages} {663} (\bibinfo {year} {2004})}\BibitemShut {NoStop}%
\bibitem [{\citenamefont {T{\"a}uber}\ \emph {et~al.}(2005)\citenamefont
  {T{\"a}uber}, \citenamefont {Howard},\ and\ \citenamefont
  {Vollmayr-Lee}}]{TauberHowardVollmayrLee05}%
  \BibitemOpen
  \bibfield  {author} {\bibinfo {author} {\bibfnamefont {U.~C.}\ \bibnamefont
  {T{\"a}uber}}, \bibinfo {author} {\bibfnamefont {M.}~\bibnamefont {Howard}},
  \ and\ \bibinfo {author} {\bibfnamefont {B.~P.}\ \bibnamefont
  {Vollmayr-Lee}},\ }\href@noop {} {\bibfield  {journal} {\bibinfo  {journal}
  {J. Phys. A}\ }\textbf {\bibinfo {volume} {38}},\ \bibinfo {pages} {R79}
  (\bibinfo {year} {2005})}\BibitemShut {NoStop}%
\bibitem [{\citenamefont {Vojta}(2006)}]{Vojta06}%
  \BibitemOpen
  \bibfield  {author} {\bibinfo {author} {\bibfnamefont {T.}~\bibnamefont
  {Vojta}},\ }\href {\doibase 10.1088/0305-4470/39/22/R01} {\bibfield
  {journal} {\bibinfo  {journal} {J. Phys. A}\ }\textbf {\bibinfo {volume}
  {39}},\ \bibinfo {pages} {R143} (\bibinfo {year} {2006})}\BibitemShut
  {NoStop}%
\bibitem [{\citenamefont {Ibrahim}\ \emph {et~al.}(2014)\citenamefont
  {Ibrahim}, \citenamefont {Barghathi},\ and\ \citenamefont
  {Vojta}}]{IbrahimBarghathiVojta14}%
  \BibitemOpen
  \bibfield  {author} {\bibinfo {author} {\bibfnamefont {A.~K.}\ \bibnamefont
  {Ibrahim}}, \bibinfo {author} {\bibfnamefont {H.}~\bibnamefont {Barghathi}},
  \ and\ \bibinfo {author} {\bibfnamefont {T.}~\bibnamefont {Vojta}},\ }\href
  {\doibase 10.1103/PhysRevE.90.042132} {\bibfield  {journal} {\bibinfo
  {journal} {Phys. Rev. E}\ }\textbf {\bibinfo {volume} {90}},\ \bibinfo
  {pages} {042132} (\bibinfo {year} {2014})}\BibitemShut {NoStop}%
\bibitem [{\citenamefont {Barghathi}\ and\ \citenamefont
  {Vojta}(2014)}]{BarghathiVojta14}%
  \BibitemOpen
  \bibfield  {author} {\bibinfo {author} {\bibfnamefont {H.}~\bibnamefont
  {Barghathi}}\ and\ \bibinfo {author} {\bibfnamefont {T.}~\bibnamefont
  {Vojta}},\ }\href {\doibase 10.1103/PhysRevLett.113.120602} {\bibfield
  {journal} {\bibinfo  {journal} {Phys. Rev. Lett.}\ }\textbf {\bibinfo
  {volume} {113}},\ \bibinfo {pages} {120602} (\bibinfo {year}
  {2014})}\BibitemShut {NoStop}%
\bibitem [{\citenamefont {Barghathi}\ and\ \citenamefont
  {Vojta}(2012)}]{BarghathiVojta12}%
  \BibitemOpen
  \bibfield  {author} {\bibinfo {author} {\bibfnamefont {H.}~\bibnamefont
  {Barghathi}}\ and\ \bibinfo {author} {\bibfnamefont {T.}~\bibnamefont
  {Vojta}},\ }\href {\doibase 10.1103/PhysRevLett.109.170603} {\bibfield
  {journal} {\bibinfo  {journal} {Phys. Rev. Lett.}\ }\textbf {\bibinfo
  {volume} {109}},\ \bibinfo {pages} {170603} (\bibinfo {year}
  {2012})}\BibitemShut {NoStop}%
\bibitem [{\citenamefont {Harris}(1974)}]{HarrisTE74}%
  \BibitemOpen
  \bibfield  {author} {\bibinfo {author} {\bibfnamefont {T.~E.}\ \bibnamefont
  {Harris}},\ }\href {\doibase doi:10.1214/aop/1176996493} {\bibfield
  {journal} {\bibinfo  {journal} {Ann. Prob.}\ }\textbf {\bibinfo {volume}
  {2}},\ \bibinfo {pages} {969} (\bibinfo {year} {1974})}\BibitemShut {NoStop}%
\bibitem [{\citenamefont {Grassberger}\ and\ \citenamefont {de~la
  Torre}(1979)}]{GrassbergerdelaTorre79}%
  \BibitemOpen
  \bibfield  {author} {\bibinfo {author} {\bibfnamefont {P.}~\bibnamefont
  {Grassberger}}\ and\ \bibinfo {author} {\bibfnamefont {A.}~\bibnamefont
  {de~la Torre}},\ }\href {\doibase 10.1016/0003-4916(79)90207-0} {\bibfield
  {journal} {\bibinfo  {journal} {Ann. Phys. (NY)}\ }\textbf {\bibinfo {volume}
  {122}},\ \bibinfo {pages} {373} (\bibinfo {year} {1979})}\BibitemShut
  {NoStop}%
\bibitem [{\citenamefont {Hinrichsen}(1997)}]{Hinrichsen97}%
  \BibitemOpen
  \bibfield  {author} {\bibinfo {author} {\bibfnamefont {H.}~\bibnamefont
  {Hinrichsen}},\ }\href@noop {} {\bibfield  {journal} {\bibinfo  {journal}
  {Phys. Rev. E}\ }\textbf {\bibinfo {volume} {55}},\ \bibinfo {pages} {219}
  (\bibinfo {year} {1997})}\BibitemShut {NoStop}%
\bibitem [{Note2()}]{Note2}%
  \BibitemOpen
  \bibinfo {note} {According to Ref.\ \cite {LeeVojta10}, the qualitative
  behavior for $\protect \mathaccentV {bar}016\mu _k \not =\mu _k$ is identical
  to that for $\protect \mathaccentV {bar}016\mu _k = \mu _k$. Moreover, the
  precise value of $\sigma $ is not important as long as it is
  nonzero.}\BibitemShut {Stop}%
\bibitem [{\citenamefont {Lee}\ and\ \citenamefont {Vojta}(2010)}]{LeeVojta10}%
  \BibitemOpen
  \bibfield  {author} {\bibinfo {author} {\bibfnamefont {M.~Y.}\ \bibnamefont
  {Lee}}\ and\ \bibinfo {author} {\bibfnamefont {T.}~\bibnamefont {Vojta}},\
  }\href@noop {} {\bibfield  {journal} {\bibinfo  {journal} {Phys. Rev. E}\
  }\textbf {\bibinfo {volume} {81}},\ \bibinfo {pages} {061128} (\bibinfo
  {year} {2010})}\BibitemShut {NoStop}%
\bibitem [{\citenamefont {Lee}\ and\ \citenamefont {Vojta}(2011)}]{LeeVojta11}%
  \BibitemOpen
  \bibfield  {author} {\bibinfo {author} {\bibfnamefont {M.~Y.}\ \bibnamefont
  {Lee}}\ and\ \bibinfo {author} {\bibfnamefont {T.}~\bibnamefont {Vojta}},\
  }\href {\doibase 10.1103/PhysRevE.83.011114} {\bibfield  {journal} {\bibinfo
  {journal} {Phys. Rev. E}\ }\textbf {\bibinfo {volume} {83}},\ \bibinfo
  {pages} {011114} (\bibinfo {year} {2011})}\BibitemShut {NoStop}%
\bibitem [{\citenamefont {Grassberger}\ \emph {et~al.}(1984)\citenamefont
  {Grassberger}, \citenamefont {Krause},\ and\ \citenamefont {von~der
  Twer}}]{GrassbergerKrauseTwer84}%
  \BibitemOpen
  \bibfield  {author} {\bibinfo {author} {\bibfnamefont {P.}~\bibnamefont
  {Grassberger}}, \bibinfo {author} {\bibfnamefont {F.}~\bibnamefont {Krause}},
  \ and\ \bibinfo {author} {\bibfnamefont {T.}~\bibnamefont {von~der Twer}},\
  }\href@noop {} {\bibfield  {journal} {\bibinfo  {journal} {J. Phys. A}\
  }\textbf {\bibinfo {volume} {17}},\ \bibinfo {pages} {L105} (\bibinfo {year}
  {1984})}\BibitemShut {NoStop}%
\bibitem [{\citenamefont {Dornic}\ \emph {et~al.}(2001)\citenamefont {Dornic},
  \citenamefont {Chat\'e}, \citenamefont {Chave},\ and\ \citenamefont
  {Hinrichsen}}]{DCCH01}%
  \BibitemOpen
  \bibfield  {author} {\bibinfo {author} {\bibfnamefont {I.}~\bibnamefont
  {Dornic}}, \bibinfo {author} {\bibfnamefont {H.}~\bibnamefont {Chat\'e}},
  \bibinfo {author} {\bibfnamefont {J.}~\bibnamefont {Chave}}, \ and\ \bibinfo
  {author} {\bibfnamefont {H.}~\bibnamefont {Hinrichsen}},\ }\href {\doibase
  10.1103/PhysRevLett.87.045701} {\bibfield  {journal} {\bibinfo  {journal}
  {Phys. Rev. Lett.}\ }\textbf {\bibinfo {volume} {87}},\ \bibinfo {pages}
  {045701} (\bibinfo {year} {2001})}\BibitemShut {NoStop}%
\bibitem [{\citenamefont {Droz}\ \emph {et~al.}(2003)\citenamefont {Droz},
  \citenamefont {Ferreira},\ and\ \citenamefont
  {Lipowski}}]{DrozFerreiraLipowski03}%
  \BibitemOpen
  \bibfield  {author} {\bibinfo {author} {\bibfnamefont {M.}~\bibnamefont
  {Droz}}, \bibinfo {author} {\bibfnamefont {A.~L.}\ \bibnamefont {Ferreira}},
  \ and\ \bibinfo {author} {\bibfnamefont {A.}~\bibnamefont {Lipowski}},\
  }\href {\doibase 10.1103/PhysRevE.67.056108} {\bibfield  {journal} {\bibinfo
  {journal} {Phys. Rev. E}\ }\textbf {\bibinfo {volume} {67}},\ \bibinfo
  {pages} {056108} (\bibinfo {year} {2003})}\BibitemShut {NoStop}%
\bibitem [{\citenamefont {Al~Hammal}\ \emph {et~al.}(2005)\citenamefont
  {Al~Hammal}, \citenamefont {Chat\'e}, \citenamefont {Dornic},\ and\
  \citenamefont {Mu\~noz}}]{ACDM05}%
  \BibitemOpen
  \bibfield  {author} {\bibinfo {author} {\bibfnamefont {O.}~\bibnamefont
  {Al~Hammal}}, \bibinfo {author} {\bibfnamefont {H.}~\bibnamefont {Chat\'e}},
  \bibinfo {author} {\bibfnamefont {I.}~\bibnamefont {Dornic}}, \ and\ \bibinfo
  {author} {\bibfnamefont {M.~A.}\ \bibnamefont {Mu\~noz}},\ }\href {\doibase
  10.1103/PhysRevLett.94.230601} {\bibfield  {journal} {\bibinfo  {journal}
  {Phys. Rev. Lett.}\ }\textbf {\bibinfo {volume} {94}},\ \bibinfo {pages}
  {230601} (\bibinfo {year} {2005})}\BibitemShut {NoStop}%
\bibitem [{\citenamefont {Solomon}(1975)}]{Solomon75}%
  \BibitemOpen
  \bibfield  {author} {\bibinfo {author} {\bibfnamefont {F.}~\bibnamefont
  {Solomon}},\ }\href@noop {} {\bibfield  {journal} {\bibinfo  {journal} {Ann.
  Prob.}\ }\textbf {\bibinfo {volume} {3}},\ \bibinfo {pages} {1} (\bibinfo
  {year} {1975})}\BibitemShut {NoStop}%
\bibitem [{\citenamefont {Kesten}\ \emph {et~al.}(1975)\citenamefont {Kesten},
  \citenamefont {Kozlov},\ and\ \citenamefont
  {Spitzer}}]{KestenKozlovSpitzer75}%
  \BibitemOpen
  \bibfield  {author} {\bibinfo {author} {\bibfnamefont {H.}~\bibnamefont
  {Kesten}}, \bibinfo {author} {\bibfnamefont {M.}~\bibnamefont {Kozlov}}, \
  and\ \bibinfo {author} {\bibfnamefont {F.}~\bibnamefont {Spitzer}},\
  }\href@noop {} {\bibfield  {journal} {\bibinfo  {journal} {Compositio Math.}\
  }\textbf {\bibinfo {volume} {30}},\ \bibinfo {pages} {145} (\bibinfo {year}
  {1975})}\BibitemShut {NoStop}%
\bibitem [{\citenamefont {Sinai}(1982)}]{Sinai82}%
  \BibitemOpen
  \bibfield  {author} {\bibinfo {author} {\bibfnamefont {Y.~G.}\ \bibnamefont
  {Sinai}},\ }\href@noop {} {\bibfield  {journal} {\bibinfo  {journal} {Theor.
  Probab. Appl.}\ }\textbf {\bibinfo {volume} {27}},\ \bibinfo {pages} {256}
  (\bibinfo {year} {1982})}\BibitemShut {NoStop}%
\bibitem [{\citenamefont {Fisher}\ \emph {et~al.}(1998)\citenamefont {Fisher},
  \citenamefont {Le~Doussal},\ and\ \citenamefont
  {Monthus}}]{FisherLeDoussalMonthus98}%
  \BibitemOpen
  \bibfield  {author} {\bibinfo {author} {\bibfnamefont {D.~S.}\ \bibnamefont
  {Fisher}}, \bibinfo {author} {\bibfnamefont {P.}~\bibnamefont {Le~Doussal}},
  \ and\ \bibinfo {author} {\bibfnamefont {C.}~\bibnamefont {Monthus}},\
  }\href@noop {} {\bibfield  {journal} {\bibinfo  {journal} {Phys. Rev. Lett}\
  }\textbf {\bibinfo {volume} {80}},\ \bibinfo {pages} {3539} (\bibinfo {year}
  {1998})}\BibitemShut {NoStop}%
\bibitem [{\citenamefont {Fisher}\ \emph {et~al.}(2001)\citenamefont {Fisher},
  \citenamefont {Le~Doussal},\ and\ \citenamefont
  {Monthus}}]{FisherLeDoussalMonthus01}%
  \BibitemOpen
  \bibfield  {author} {\bibinfo {author} {\bibfnamefont {D.~S.}\ \bibnamefont
  {Fisher}}, \bibinfo {author} {\bibfnamefont {P.}~\bibnamefont {Le~Doussal}},
  \ and\ \bibinfo {author} {\bibfnamefont {C.}~\bibnamefont {Monthus}},\
  }\href@noop {} {\bibfield  {journal} {\bibinfo  {journal} {Phys. Rev. E}\
  }\textbf {\bibinfo {volume} {64}},\ \bibinfo {pages} {066107} (\bibinfo
  {year} {2001})}\BibitemShut {NoStop}%
\bibitem [{\citenamefont {Doussal}\ \emph {et~al.}(1999)\citenamefont
  {Doussal}, \citenamefont {Monthus},\ and\ \citenamefont
  {Fisher}}]{LeDoussalMonthusFisher99}%
  \BibitemOpen
  \bibfield  {author} {\bibinfo {author} {\bibfnamefont {P.~L.}\ \bibnamefont
  {Doussal}}, \bibinfo {author} {\bibfnamefont {C.}~\bibnamefont {Monthus}}, \
  and\ \bibinfo {author} {\bibfnamefont {D.~S.}\ \bibnamefont {Fisher}},\
  }\href@noop {} {\bibfield  {journal} {\bibinfo  {journal} {Phys. Rev. E}\
  }\textbf {\bibinfo {volume} {59}},\ \bibinfo {pages} {4795} (\bibinfo {year}
  {1999})}\BibitemShut {NoStop}%
\bibitem [{\citenamefont {Grinstein}\ and\ \citenamefont
  {Fernandez}(1984)}]{Grinstein84}%
  \BibitemOpen
  \bibfield  {author} {\bibinfo {author} {\bibfnamefont {G.}~\bibnamefont
  {Grinstein}}\ and\ \bibinfo {author} {\bibfnamefont {J.~F.}\ \bibnamefont
  {Fernandez}},\ }\href {\doibase 10.1103/PhysRevB.29.6389} {\bibfield
  {journal} {\bibinfo  {journal} {Phys. Rev. B}\ }\textbf {\bibinfo {volume}
  {29}},\ \bibinfo {pages} {6389} (\bibinfo {year} {1984})}\BibitemShut
  {NoStop}%
\bibitem [{\citenamefont {Hooyberghs}\ \emph {et~al.}(2003)\citenamefont
  {Hooyberghs}, \citenamefont {Igl\'oi},\ and\ \citenamefont
  {Vanderzande}}]{HooyberghsIgloiVanderzande03}%
  \BibitemOpen
  \bibfield  {author} {\bibinfo {author} {\bibfnamefont {J.}~\bibnamefont
  {Hooyberghs}}, \bibinfo {author} {\bibfnamefont {F.}~\bibnamefont {Igl\'oi}},
  \ and\ \bibinfo {author} {\bibfnamefont {C.}~\bibnamefont {Vanderzande}},\
  }\href {\doibase 10.1103/PhysRevLett.90.100601} {\bibfield  {journal}
  {\bibinfo  {journal} {Phys. Rev. Lett.}\ }\textbf {\bibinfo {volume} {90}},\
  \bibinfo {pages} {100601} (\bibinfo {year} {2003})}\BibitemShut {NoStop}%
\bibitem [{\citenamefont {Vojta}\ and\ \citenamefont
  {Dickison}(2005)}]{VojtaDickison05}%
  \BibitemOpen
  \bibfield  {author} {\bibinfo {author} {\bibfnamefont {T.}~\bibnamefont
  {Vojta}}\ and\ \bibinfo {author} {\bibfnamefont {M.}~\bibnamefont
  {Dickison}},\ }\href {\doibase 10.1103/PhysRevE.72.036126} {\bibfield
  {journal} {\bibinfo  {journal} {Phys. Rev. E}\ }\textbf {\bibinfo {volume}
  {72}},\ \bibinfo {pages} {036126} (\bibinfo {year} {2005})}\BibitemShut
  {NoStop}%
\bibitem [{\citenamefont {Vojta}\ \emph {et~al.}(2009)\citenamefont {Vojta},
  \citenamefont {Farquhar},\ and\ \citenamefont {Mast}}]{VojtaFarquharMast09}%
  \BibitemOpen
  \bibfield  {author} {\bibinfo {author} {\bibfnamefont {T.}~\bibnamefont
  {Vojta}}, \bibinfo {author} {\bibfnamefont {A.}~\bibnamefont {Farquhar}}, \
  and\ \bibinfo {author} {\bibfnamefont {J.}~\bibnamefont {Mast}},\ }\href
  {\doibase 10.1103/PhysRevE.79.011111} {\bibfield  {journal} {\bibinfo
  {journal} {Phys. Rev. E}\ }\textbf {\bibinfo {volume} {79}},\ \bibinfo
  {pages} {011111} (\bibinfo {year} {2009})}\BibitemShut {NoStop}%
\bibitem [{\citenamefont {Vojta}(2012)}]{Vojta12}%
  \BibitemOpen
  \bibfield  {author} {\bibinfo {author} {\bibfnamefont {T.}~\bibnamefont
  {Vojta}},\ }\href {\doibase 10.1103/PhysRevE.86.051137} {\bibfield  {journal}
  {\bibinfo  {journal} {Phys. Rev. E}\ }\textbf {\bibinfo {volume} {86}},\
  \bibinfo {pages} {051137} (\bibinfo {year} {2012})}\BibitemShut {NoStop}%
\bibitem [{\citenamefont {Pigolotti}\ and\ \citenamefont
  {Cencini}(2010)}]{PigolottiCencini10}%
  \BibitemOpen
  \bibfield  {author} {\bibinfo {author} {\bibfnamefont {S.}~\bibnamefont
  {Pigolotti}}\ and\ \bibinfo {author} {\bibfnamefont {M.}~\bibnamefont
  {Cencini}},\ }\href@noop {} {\bibfield  {journal} {\bibinfo  {journal} {J.
  Theor. Biology}\ }\textbf {\bibinfo {volume} {265}},\ \bibinfo {pages} {609}
  (\bibinfo {year} {2010})}\BibitemShut {NoStop}%
\bibitem [{\citenamefont {Borile}\ \emph {et~al.}(2013)\citenamefont {Borile},
  \citenamefont {Maritan},\ and\ \citenamefont {Mu\~noz}}]{BorileMaritanMunoz}%
  \BibitemOpen
  \bibfield  {author} {\bibinfo {author} {\bibfnamefont {C.}~\bibnamefont
  {Borile}}, \bibinfo {author} {\bibfnamefont {A.}~\bibnamefont {Maritan}}, \
  and\ \bibinfo {author} {\bibfnamefont {M.~A.}\ \bibnamefont {Mu\~noz}},\
  }\href {http://stacks.iop.org/1742-5468/2013/i=04/a=P04032} {\bibfield
  {journal} {\bibinfo  {journal} {Journal of Statistical Mechanics: Theory and
  Experiment}\ }\textbf {\bibinfo {volume} {2013}},\ \bibinfo {pages} {P04032}
  (\bibinfo {year} {2013})}\BibitemShut {NoStop}%
\bibitem [{\citenamefont {Villa~Mart\'{\i}n}\ \emph {et~al.}(2014)\citenamefont
  {Villa~Mart\'{\i}n}, \citenamefont {Bonachela},\ and\ \citenamefont
  {Mu\~noz}}]{MartnBonachelaMunoz}%
  \BibitemOpen
  \bibfield  {author} {\bibinfo {author} {\bibfnamefont {P.}~\bibnamefont
  {Villa~Mart\'{\i}n}}, \bibinfo {author} {\bibfnamefont {J.~A.}\ \bibnamefont
  {Bonachela}}, \ and\ \bibinfo {author} {\bibfnamefont {M.~A.}\ \bibnamefont
  {Mu\~noz}},\ }\href {\doibase 10.1103/PhysRevE.89.012145} {\bibfield
  {journal} {\bibinfo  {journal} {Phys. Rev. E}\ }\textbf {\bibinfo {volume}
  {89}},\ \bibinfo {pages} {012145} (\bibinfo {year} {2014})}\BibitemShut
  {NoStop}%
\bibitem [{Note3()}]{Note3}%
  \BibitemOpen
  \bibinfo {note} {In dimensions $d>2$ the number of different configurations
  of nearest-neighbors in states $I_1$ vs $I_2$ increases. As a result the
  number of nontrivial hopping rates ratios is more than one and increases with
  the dimensionality of the system. In the Ising model there is only one
  parameter to fine tune map these ratios, specifically, the ratio
  $J/T$.}\BibitemShut {Stop}%
\bibitem [{\citenamefont {Ben-Naim}\ \emph {et~al.}(1996)\citenamefont
  {Ben-Naim}, \citenamefont {Frachebourg},\ and\ \citenamefont
  {Krapivsky}}]{Ben-NaimFrachebourgKrapivsky1996}%
  \BibitemOpen
  \bibfield  {author} {\bibinfo {author} {\bibfnamefont {E.}~\bibnamefont
  {Ben-Naim}}, \bibinfo {author} {\bibfnamefont {L.}~\bibnamefont
  {Frachebourg}}, \ and\ \bibinfo {author} {\bibfnamefont {P.~L.}\ \bibnamefont
  {Krapivsky}},\ }\href {\doibase 10.1103/PhysRevE.53.3078} {\bibfield
  {journal} {\bibinfo  {journal} {Phys. Rev. E}\ }\textbf {\bibinfo {volume}
  {53}},\ \bibinfo {pages} {3078} (\bibinfo {year} {1996})}\BibitemShut
  {NoStop}%
\bibitem [{\citenamefont {Hinrichsen}(2000{\natexlab{b}})}]{Hinrichsen00b}%
  \BibitemOpen
  \bibfield  {author} {\bibinfo {author} {\bibfnamefont {H.}~\bibnamefont
  {Hinrichsen}},\ }\href@noop {} {\bibfield  {journal} {\bibinfo  {journal}
  {Braz. J. Phys.}\ }\textbf {\bibinfo {volume} {30}},\ \bibinfo {pages} {69}
  (\bibinfo {year} {2000}{\natexlab{b}})}\BibitemShut {NoStop}%
\bibitem [{\citenamefont {Corte}\ \emph {et~al.}(2008)\citenamefont {Corte},
  \citenamefont {Chaikin}, \citenamefont {Gollub},\ and\ \citenamefont
  {Pine}}]{CCGP08}%
  \BibitemOpen
  \bibfield  {author} {\bibinfo {author} {\bibfnamefont {L.}~\bibnamefont
  {Corte}}, \bibinfo {author} {\bibfnamefont {P.~M.}\ \bibnamefont {Chaikin}},
  \bibinfo {author} {\bibfnamefont {J.~P.}\ \bibnamefont {Gollub}}, \ and\
  \bibinfo {author} {\bibfnamefont {D.~J.}\ \bibnamefont {Pine}},\ }\href@noop
  {} {\bibfield  {journal} {\bibinfo  {journal} {Nature Physics}\ }\textbf
  {\bibinfo {volume} {4}},\ \bibinfo {pages} {420} (\bibinfo {year}
  {2008})}\BibitemShut {NoStop}%
\bibitem [{\citenamefont {Franceschini}\ \emph {et~al.}(2011)\citenamefont
  {Franceschini}, \citenamefont {Filippidi}, \citenamefont {Guazzelli},\ and\
  \citenamefont {Pine}}]{FFGP11}%
  \BibitemOpen
  \bibfield  {author} {\bibinfo {author} {\bibfnamefont {A.}~\bibnamefont
  {Franceschini}}, \bibinfo {author} {\bibfnamefont {E.}~\bibnamefont
  {Filippidi}}, \bibinfo {author} {\bibfnamefont {E.}~\bibnamefont
  {Guazzelli}}, \ and\ \bibinfo {author} {\bibfnamefont {D.~J.}\ \bibnamefont
  {Pine}},\ }\href {\doibase 10.1103/PhysRevLett.107.250603} {\bibfield
  {journal} {\bibinfo  {journal} {Phys. Rev. Lett.}\ }\textbf {\bibinfo
  {volume} {107}},\ \bibinfo {pages} {250603} (\bibinfo {year}
  {2011})}\BibitemShut {NoStop}%
\bibitem [{\citenamefont {Takeuchi}\ \emph {et~al.}(2007)\citenamefont
  {Takeuchi}, \citenamefont {Kuroda}, \citenamefont {Chate},\ and\
  \citenamefont {Sano}}]{TKCS07}%
  \BibitemOpen
  \bibfield  {author} {\bibinfo {author} {\bibfnamefont {K.~A.}\ \bibnamefont
  {Takeuchi}}, \bibinfo {author} {\bibfnamefont {M.}~\bibnamefont {Kuroda}},
  \bibinfo {author} {\bibfnamefont {H.}~\bibnamefont {Chate}}, \ and\ \bibinfo
  {author} {\bibfnamefont {M.}~\bibnamefont {Sano}},\ }\href@noop {} {\bibfield
   {journal} {\bibinfo  {journal} {Phys. Rev. Lett.}\ }\textbf {\bibinfo
  {volume} {99}},\ \bibinfo {pages} {234503} (\bibinfo {year}
  {2007})}\BibitemShut {NoStop}%
\bibitem [{\citenamefont {Okuma}\ \emph {et~al.}(2011)\citenamefont {Okuma},
  \citenamefont {Tsugawa},\ and\ \citenamefont
  {Motohashi}}]{OkumaTsugawaMotohashi11}%
  \BibitemOpen
  \bibfield  {author} {\bibinfo {author} {\bibfnamefont {S.}~\bibnamefont
  {Okuma}}, \bibinfo {author} {\bibfnamefont {Y.}~\bibnamefont {Tsugawa}}, \
  and\ \bibinfo {author} {\bibfnamefont {A.}~\bibnamefont {Motohashi}},\ }\href
  {\doibase 10.1103/PhysRevB.83.012503} {\bibfield  {journal} {\bibinfo
  {journal} {Phys. Rev. B}\ }\textbf {\bibinfo {volume} {83}},\ \bibinfo
  {pages} {012503} (\bibinfo {year} {2011})}\BibitemShut {NoStop}%
\bibitem [{\citenamefont {Korolev}\ \emph {et~al.}(2011)\citenamefont
  {Korolev}, \citenamefont {Xavier}, \citenamefont {Nelson},\ and\
  \citenamefont {Foster}}]{KXNF11}%
  \BibitemOpen
  \bibfield  {author} {\bibinfo {author} {\bibfnamefont {K.~S.}\ \bibnamefont
  {Korolev}}, \bibinfo {author} {\bibfnamefont {J.~B.}\ \bibnamefont {Xavier}},
  \bibinfo {author} {\bibfnamefont {D.~R.}\ \bibnamefont {Nelson}}, \ and\
  \bibinfo {author} {\bibfnamefont {K.~R.}\ \bibnamefont {Foster}},\
  }\href@noop {} {\bibfield  {journal} {\bibinfo  {journal} {The American
  Naturalist}\ }\textbf {\bibinfo {volume} {178}},\ \bibinfo {pages} {538}
  (\bibinfo {year} {2011})}\BibitemShut {NoStop}%
\bibitem [{\citenamefont {Korolev}\ and\ \citenamefont
  {Nelson}(2011)}]{KorolevNelson11}%
  \BibitemOpen
  \bibfield  {author} {\bibinfo {author} {\bibfnamefont {K.~S.}\ \bibnamefont
  {Korolev}}\ and\ \bibinfo {author} {\bibfnamefont {D.~R.}\ \bibnamefont
  {Nelson}},\ }\href {\doibase 10.1103/PhysRevLett.107.088103} {\bibfield
  {journal} {\bibinfo  {journal} {Phys. Rev. Lett.}\ }\textbf {\bibinfo
  {volume} {107}},\ \bibinfo {pages} {088103} (\bibinfo {year}
  {2011})}\BibitemShut {NoStop}%
\bibitem [{\citenamefont {Villain}(1984)}]{Villain84}%
  \BibitemOpen
  \bibfield  {author} {\bibinfo {author} {\bibfnamefont {J.}~\bibnamefont
  {Villain}},\ }\href {\doibase 10.1103/PhysRevLett.52.1543} {\bibfield
  {journal} {\bibinfo  {journal} {Phys. Rev. Lett.}\ }\textbf {\bibinfo
  {volume} {52}},\ \bibinfo {pages} {1543} (\bibinfo {year}
  {1984})}\BibitemShut {NoStop}%
\bibitem [{\citenamefont {Grinstein}\ and\ \citenamefont
  {Ma}(1983)}]{Grinstein83}%
  \BibitemOpen
  \bibfield  {author} {\bibinfo {author} {\bibfnamefont {G.}~\bibnamefont
  {Grinstein}}\ and\ \bibinfo {author} {\bibfnamefont {S.-k.}\ \bibnamefont
  {Ma}},\ }\href {\doibase 10.1103/PhysRevB.28.2588} {\bibfield  {journal}
  {\bibinfo  {journal} {Phys. Rev. B}\ }\textbf {\bibinfo {volume} {28}},\
  \bibinfo {pages} {2588} (\bibinfo {year} {1983})}\BibitemShut {NoStop}%
\end{thebibliography}%
\end{document}